\documentclass[10pt]{article}
\usepackage[a4paper,centering,hscale=0.7,vscale=0.75]{geometry}

\usepackage{lmodern}
\usepackage[T1]{fontenc}

\usepackage{mathtools}      
\usepackage{amsthm,amssymb} 
\usepackage{mathrsfs}
\usepackage{xcolor}
\usepackage{booktabs}

\usepackage[colorlinks=true]{hyperref}

\usepackage{pgfplots}
\pgfplotsset{width=10cm,compat=newest}

\newtheorem{prop}{Proposition}[section]

\newtheorem{lemma}[prop]{Lemma}
\newtheorem{coroll}[prop]{Corollary}

\theoremstyle{remark}
\newtheorem{rmk}[prop]{Remark}

\newcommand{\bC}{\mathbb{C}}

\newcommand{\cF}{\mathscr{F}}
\newcommand{\cL}{\mathscr{L}}
\newcommand{\bN}{\mathbb{N}}
\renewcommand{\P}{\mathbb{P}}
\newcommand{\bQ}{\mathbb{Q}}
\newcommand{\Q}{\mathop{{}\mathbb{Q}}\nolimits}

\newcommand{\erre}{\mathbb{R}}

\renewcommand{\geq}{\geqslant}
\renewcommand{\leq}{\leqslant}

\numberwithin{equation}{section}

\DeclarePairedDelimiter\abs{\lvert}{\rvert}
\DeclarePairedDelimiter\norm{\lVert}{\rVert}

\DeclarePairedDelimiterX\ip[2]{\langle}{\rangle}{#1,#2}

\mathtoolsset{centercolon}

\makeatletter
\renewenvironment{enumerate}{%
  \ifnum \@enumdepth >3 \@toodeep\else
      \advance\@enumdepth \@ne
      \edef\@enumctr{enum\romannumeral\the\@enumdepth}\list
      {\csname label\@enumctr\endcsname}{\usecounter
        {\@enumctr}\def\makelabel##1{\hss\llap{\upshape##1}}}\fi
}{%
  \endlist
}

\renewenvironment{itemize}{%
  \ifnum\@itemdepth>3 \@toodeep
  \else \advance\@itemdepth\@ne
    \edef\@itemitem{labelitem\romannumeral\the\@itemdepth}%
    \list{\csname\@itemitem\endcsname}%
      {\def\makelabel##1{\hss\llap{\upshape##1}}}%
  \fi
}{%
  \endlist
}

\makeatother

\begin{document}

\title{On the relative performance of some parametric and
  nonparametric estimators of option prices}
\author{Carlo Marinelli\thanks{Department of Mathematics, University
    College London, Gower Street, London WC1E 6BT, UK.} \ and Stefano
  D'Addona\thanks{Dipartimento di Scienze Politiche, Università degli
    studi Roma Tre, via G.~Chiabrera 199, 00145 Roma, Italy,}}
\date{\normalsize November 28, 2024}

\maketitle

\begin{abstract}
  We examine the empirical performance of some parametric and
  nonparametric estimators of prices of options with a fixed time to
  maturity, focusing on variance-gamma and Heston models on one side,
  and on expansions in Hermite functions on the other side. The latter
  class of estimators can be seen as perturbations of the classical
  Black-Scholes model. The comparison between parametric and
  Hermite-based models having the same ``degrees of freedom'' is
  emphasized.  The main criterion is the out-of-sample relative
  pricing error on a dataset of historical option prices on the
  S\&P500 index. Prior to the main empirical study, the approximation
  of variance-gamma and Heston densities by series of Hermite
  functions is studied, providing explicit expressions for the
  coefficients of the expansion in the former case, and integral
  expressions involving the explicit characteristic function in the
  latter case. Moreover, these approximations are investigated
  numerically on a few test cases, indicating that expansions in
  Hermite functions with few terms achieve competitive accuracy in the
  estimation of Heston densities and the pricing of (European)
  options, but they perform less effectively with variance-gamma
  densities. On the other hand, the main large-scale empirical study
  show that parsimonious Hermite estimators can even outperform the
  Heston model in terms of pricing errors.  These results underscore
  the trade-offs inherent in model selection and calibration, and
  their empirical fit in practical applications.
\end{abstract}


\section{Introduction}
Our main goal is to test the empirical performance in the estimation
of option prices of some well-known parametric model for the density
of logarithmic asset prices versus nonparametric approximations of the
same density by means of expansions in (shifted and rescaled) Hermite
functions. We are particularly interested in comparisons among
parametric and nonparametric models having, in a certain sense, the
same degrees of freedom. In fact, if the density to be approximated is
square-integrable, i.e. it belongs to \(L^2:=L^2(\erre)\), then it
admits an expansion in Hermite functions, a truncation of which
depends only on a finite number of real coefficients. This
approximation is nonparametric in the sense that it does not assume
that the density belongs to any parametric class.
As benchmark parametric models for the density of logarithmic returns
we consider the variance-gamma model, as defined in \cite{MCC}, and
the Heston model (see \cite{Heston}), the former of which is specified
by three parameters and the latter by five (see
\S\S\ref{sec:VG}-\ref{sec:Heston} for details).
Denoting the density of logarithmic returns at a fixed time \(t\) by
\(f\), and assuming that \(f\) belongs to \(L^2\), we shall consider
approximations of \(f\) of the type
\[
f_n(x) := \sum_{k=0}^n \alpha_k h_k\biggl( \frac{x-b}{a} \biggr),
\]
where \(h_k\) is the Hermite functions of order \(k\) defined by
\(h_k(x) = H_k(\sqrt{2}x) \exp (-x^2/2)\), with \(H_k\) the Hermite
polynomial of degree \(k\), the coefficients \((\alpha_k)\) are real
numbers, and the constants \(a,b\) are scale and location parameters,
respectively (see \S\ref{sec:He} for details). For every integer
\(n\), a function \(f_n\) is determined by \(n+3\) parameters, to wit,
the \(n+1\) coefficients \((\alpha_k)\) and the scale and location
parameters \(a\) and \(b\). This class of approximations and some of
its properties were studied in \cite{cm:Herm}, where it was observed
that \(f_n\) can be seen as a perturbation of a Gaussian density
(corresponding to \(n=0\)), or, in other words, of the classical
Black-Scholes model. Assuming that the reference probability measure
is a pricing measure, it was also observed that setting \(b=-a^2/2\),
as is necessary in the limiting case \(n=0\), does not significantly
alter the empirical accuracy of the approximation. Therefore we shall
mostly restrict to this specific setting, in which case the
approximation \(f_{n,p}\), defined as \(f_n\) with \(b=-a^2/2\), is
uniquely determined by \(n+2\) parameters. In general neither \(f_n\)
nor \(f_{n,p}\) are themselves densities, as they may assume negative
values on nonempty subsets of \(\erre\) and do not necessarily
integrate to one. Similarly, assuming for simplicity (and without loss
of generality) that the asset price is equal to one at time zero, the
function \(x \mapsto e^x f(x)\) integrates to one, but the same cannot
be expected to hold replacing \(f\) by \(f_n\) or \(f_{n,p}\). It
turns out that the two constraints on approximations of \(f\) to
integrate to one and to integrate to one when multiplied by the
exponential function are independent and reduce the degrees of freedom
by two. A Hermite approximation of \(f\) satisfying these two
constraints will be said to have the approximate martingale
property. An approximation of the type \(f_{n,p}\) with the
approximate martingale property, that will be denoted by \(f_{n,m}\),
is then uniquely determined by \(n\) parameters. Natural matches for
parametric models of the density depending on \(n\) parameters are
then \(f_{n-3}\), \(f_{n-2,p}\), and \(f_{n,m}\), whenever these are
well defined. In particular, we shall test, on the basis of several
criteria, the variance-gamma model against \(f_{1,p}\) and
\(f_{3,m}\), and the Heston model against \(f_2\), \(f_{3,p}\) and
\(f_{5,m}\).
This can also be seen as an attempt to understand whether a
general-purpose approximation scheme can reproduce, with reasonable
accuracy, ``informed'' parametric models especially designed for
financial asset returns.

The expirical evidence that we obtain from numerical experiments on
simulated and real data is somewhat mixed. The density of a
variance-gamma process, that, incidentally, does not belong to \(L^2\)
at small times, appears to be harder to approximate with a good level
of precision than the density of a Heston process, both in terms of
the \(L^2\) norm of the difference between the target density and the
approximating one and in terms of the pricing error. More precisely,
option prices implied by a Heston density can be approximated quite
well by Hermite estimators, while the results are poorer for option
prices implied by a variance-gamma density, even though some
estimators perform well in particular cases. In an extensive empirical
study on real data, where all estimators are compared on the basis of
their out-of-sample pricing error, it turns out, as expected, that the
richer (in terms of number of parameters) Heston model performs better
than the variance-gamma model, but, rather surprisingly, a simple
approximation of the implicit density of logarithmic returns of type
\(f_2\), hence a linear combination of very few scaled and shifted
Hermite functions, outperforms also the Heston model.

Another motivation for the empirical study conducted here is to
complement the results of our previous work \cite{cm:Herm}, where
several Hermite-based nonparametric option price estimators were
compared to each other as well as to an elementary nonparametric
method relying on linear interpolation on the implied volatility
curve, but no comparison with nontrivial well-established parametric
models was carried out. Even though \cite{cm:Herm} considered more
Hermite-based approximations than those used here, results are
directly and easily comparable with those reported therein.

While we only deal with approximations of densities of logarithmic
returns at fixed times, it would of course be interesting to study the
(presumably much harder) problem of approximating joint densities at
multiple times by nonparametric models based on Hermite functions. The
closely related problem of estimating prices of options with different
maturities by a unique time-dependent Hermite-based model is of course
equally interesting, especially in view of the well-known fact that
the Heston model performs reasonably well in the approximation of
prices of options with medium-long maturity.

Some literature on financial applications of Hermite functions, as
well as of other basis functions involving orthogonal polynomials, is
cited in \cite{cm:Herm}, to which we refer. Other examples are
\cite{Ack:Ortho} and the very recent \cite{Zhao:Poly}, that also
contains references to applications of the related Gram–Charlier type
A series. The type of problems treated in all these references,
however, is different from ours. Applications of Hermite polynomials
to some classical statistical problems and relevant pointers to the
literature can be found, e.g., in \cite{SteVa:Herm}.

We conclude this introductory section with a brief outline of the rest
of the text: after fixing some basic notation and assumptions in
\S\ref{sec:prel}, we recall the definition of the Hermite option price
estimator in \S\ref{sec:He}, referring to \cite{cm:Herm} for more
details. Sections \ref{sec:VG} and \ref{sec:Heston} contain brief
presentations of the variance-gamma and Heston processes, including
sufficient conditions for their densities to belong to \(L^2\), and of
the corresponding option pricing formulas used throughout. Some
theoretical aspects of the expansions in series of Hermite functions
of logarithmic returns' densities are studied in \S\ref{sec:Had}. In
particular, the smoothness of the objective function arising in the
optimization of \(f_n\) with respect to the parameters \(a,b\) is
proved and its first- and second-order total derivatives are
computed. Morover, the coefficients of a constrained Hermite expansion
in terms of those of an unconstrained expansion are characterized
explicitly. Furthermore, it is shown that, in the variance-gamma case,
the coefficients of the Hermite expansion of a density can be computed
in terms of integrals with respect to gamma measures, while, in the
Heston case, they can be obtained by the Hermite expansion of an
explicit function involving the Fourier transform of the density,
which is known in closed form.
In the final section \ref{sec:res}, results of numerical computations
on synthetic and real data are reported and discussed. We first
consider two test cases, where a variance-gamma density and a Heston
density, as well as a set of corresponding option prices, are
estimated by the Hermite approximation schemes described above. Then
we provide a comparison of the empirical performance of all the
parametric models and Hermite approximations on S\&P500 option data,
by analyzing the respective out-of-sample pricing errors.


\section{Preliminaries}
\label{sec:prel}

\subsection{Notation and basic assumptions}
We shall use the following notation:
\(\erre_+ := \mathopen[0,+\infty\mathclose[\),
\(\erre_+^\times := \erre_+ \setminus \{0\}\), and \(A \lesssim B\) to
mean that there exists a constant \(c\) such that \(A \leq cB\), with
subscripts indicating parameters on which \(c\) depends. Moreover,
\(A \eqsim B\) stands for \(A \lesssim B\) and \(B \lesssim A\).
Numbers will be reported with three significant digits, with the
exception of Tables \ref{tab:BS-Hes-VG} and \ref{tab:Hermite} (for
formatting reasons).
The spaces \(L^p(\erre)\), \(p \in [0,\infty]\), are denoted simply by
\(L^p\). The scalar product of two functions \(f,g \in L^2\) is
denoted by \(\ip{f}{g}\).
The distribution function and the density function of the standard
Gaussian measure on \(\erre\) are denoted by \(\Phi\) and \(\phi\),
respectively. The few special functions used in the text are defined
as in \cite{DLMF}.

Let \(T \in \erre_+^\times\) and \((\Omega,\cF,\P)\) be a probability
space endowed with a right-continuous filtration
\((\cF_t)_{t\in[0,T]}\) such that \(\cF_0\) contains the negligible
sets of \(\cF_T\). All random variables and stochastic processes will
be defined on this filtered probability space. The expectation (with
respect to \(\P\)) of a random variable \(X\) will be denoted by
\(\P X\).

Regarding asset price processes, independently of the setting
considered in the following sections, we shall use the following
conventions: \(\widehat{S}\) stands for the price process of a
dividend-paying asset, and \(\beta\) for the price process of a money
market account used as num\'eraire, about which we assume that
\[
  \beta_t =: \exp\biggl( \int_0^t r_u\,du \biggr),
\]
where \(r\), the short rate process, is positive and adapted.  The
discounted asset price process is then defined by
\(S := \widehat{S}/\beta\).
The cumulative dividend process is assumed, for simplicity, to be
determined by a constant non-random dividend rate \(q\), and that the
process \((S_te^{qt})_{t \geq 0}\) is a martingale with a respect to a
probability measure \(\bQ\), equivalent to \(\P\), that is used as
pricing measure.

The function \(\mathsf{BS}\colon \erre^6_+ \to \erre_+\) is defined,
for every \((s_0,k,r,q,\sigma,t)\), as the Black-Scholes price at time
zero of a put option with strike \(k\) and expiration time \(t\) on an
asset with price \(s\) at time zero, volatility \(\sigma\), and
dividend rate \(q\), assuming a constant short rate equal to \(r\).

For later reference, we recall a convergence result proved in
\cite{cm:repr}: for a function \(f\in L^1 \cap L^2\), let
\(\pi\colon \erre_+ \to \erre_+\) be defined by
\[
\pi(k) := \int_\erre {(k-e^x)}^+ f(x)\,dx.
\]
If \((f_n)\) is a sequence of functions in \(L^1 \cap L^2\) converging
to \(f\) in \(L^2\), then, defining \(\pi_n\) in the obvious way, one
has, for any \(k_0 \in \erre_+^\times\),
\[
  \pi(k) = \frac{k}{k_0} \pi(k_0)
  + \lim_{n \to \infty} \Bigl( \pi^n(k) - \frac{k}{k_0} \pi^n(k_0) \Bigr).
\]
For a given \(n \in \bN\), in the approximation
\begin{equation}
  \label{eq:pin}
  \begin{split}
    \pi(k)
    &\approx \frac{k}{k_0} \pi(k_0) + \pi^n(k) - \frac{k}{k_0} \pi^n(k_0)\\
    &= \pi^n(k) + \frac{k}{k_0} \bigl( \pi(k_0) - \pi^n(k_0) \bigr)
  \end{split}
\end{equation}
we shall say that the right-hand side is a ``corrected'' version of
\(\pi(k)\) once \(\pi(k_0)\) is known.


\section{Hermite estimator}
\label{sec:He}
The Hermite estimator of option prices is constructed approximating
the density of logarithmic returns by a finite linear combination of
Hermite functions. The procedure is discussed at length in
\cite{cm:Herm}, to which we refer for details, while here we only
recall some essential aspects. We assume \(S_0=1\) and \(q=0\)
throughout this section for simplicity. This normalization comes at no
loss of generality, as the general case can be recovered simply
replacing \(S_t\) with \(S_0 S_t e^{qt}\) everywhere. Furthermore, let
\(t \in \erre_+^\times\) be a fixed time horizon, and assume that
\(\Q(S_t>0)=1\) and that the law of \(\log S_t\) with respect to \(\bQ\)
is absolutely continuous with a density belonging to \(L^2\).

Given constants \(a \in \erre_+^\times\) and \(b \in \erre\), define
the random variable \(X\) by \(aX + b = \log S_t\). The density \(f\)
of the random variable \(X\), that also belongs to \(L^2\), is then
approximated by a finite linear combination of Hermite functions, the
definition of which is recalled next.  The Hermite polynomials
\((H_n)_{n \geq 0}\) are defined by
\[
  H_n(x) := (-1)^n e^{x^2/2} \frac{d^n}{dx^n} e^{-x^2/2},
  \qquad n \in \mathbb{N}.
\]
For instance, the first four of them are
\[
  H_0(x) = 1, \qquad H_1(x) = x, \qquad H_2(x) = x^2-1,
  \qquad H_3(x) = x^3-3x.
\]
The Hermite functions \((h_n)_{n \geq 0}\) are defined by
\[
h_n(x) := H_n(\sqrt{2}x) e^{-x^2/2}.
\]
Using well-known properties of Hermite polynomials, it is not hard to
show that \({\norm{h_n}}_{L^2}^2 = n! \sqrt{\pi} =: c_n\) for
every \(n \in \bN\) and that \((h_n/c_n)_{n \in \bN}\) is a
complete orthonormal basis of \(L^2\). Therefore, setting, for every
\(n \in \bN\),
\[
  \alpha_n := \frac{1}{n!\sqrt{\pi}} \ip{f}{h_n} =
  \frac{1}{n!\sqrt{\pi}} \int_\erre f(x) h_n(x)\,dx
\]
and
\[
  f_n = \sum_{k=0}^n \alpha_k h_k,
\]
one has \(\lim_{n \to \infty} f_n = f\) in \(L^2\).

If \(g\colon \erre_+ \to \erre\) is a measurable function such that
\(x \mapsto g(e^{ax+b}) f(x) \in L^1\), or, equivalently, such that
\(\Q g(S_t)\) is finite, then the Hermite estimator of order \(n\) of
\[
  \Q g(S_t) = \int_\erre g(e^{ax+b}) f(x)\,dx
\]
is
\[
  \int_\erre g(e^{ax+b}) f_n(x)\,dx = \sum_{k=0}^n \alpha_k
  \int_\erre g(e^{ax+b}) h_k(x)\,dx.
\]
\begin{rmk}
  Since \(f \in L^1 \cap L^2\) and the dual of \(L^1 \cap L^2\) is
  \(L^2 + L^\infty\), the assumption
  \(x \mapsto g(e^{ax+b}) \in L^2 + L^\infty\) implies that
  \(\Q g(S_t)\) is finite. However, this assumption is not sharp, as
  \(g\colon x \mapsto x\) does not satisfy it, but \(\Q S_t=1\). The
  latter observation implies that Hermite estimators of call and put
  options are well defined.
\end{rmk}
The price at time zero of a put option with strike \(k\) is given by
\(\Q{(\beta_t^{-1} k - S_t)}^+\), which simplifies, assuming that \(r\)
is nonrandom, to
\[
  \Q\bigl( ke^{-rt} - S_t \bigr)^+ = \int_\erre
  \bigl( ke^{-rt} - e^{ax+b} \bigr)^+ f(x)\,dx,
\]
hence, setting
\[
\zeta_+ := \frac{1}{a}\bigl( \log k - rt - b \bigr),
\]
the Hermite estimator of the put option price is
\[
  k e^{-rt} \sum_{k=0}^n \alpha_k \int_{-\infty}^{\zeta_+} h_k(x)\,dx
  - \sum_{k=0}^n \alpha_k \int_{-\infty}^{\zeta_+} e^{ax+b} h_k(x)\,dx,
\]
where both integrals can be computed explicitly in terms of the
standard Gaussian cumulative distribution function \(\Phi\), or in
terms of the \href{https://dlmf.nist.gov/8}{incomplete gamma function}
(see \cite{cm:Herm} for details).

Once the Hermite estimators of put option prices have been defined,
the coefficients \((\alpha_k)\) can be calibrated, for given constants
\(a,b\), to a set of observed put option prices in a very efficient
way due to the linearity of the approximating price with respect to
\((\alpha_k)\). As a second step, calibration of the parameters
\(a,b\) can also be performed in a numerically efficient way. Several
calibration criteria and algorithms are discussed and tested
empirically in \cite{cm:Herm}.
Calibration is meant here and throughout the paper as minimization of
a loss function applied to the absolute relative pricing error,
defined as follows: if \(\pi \in \erre^N\) is a vector of observed
prices of options with a fixed time to maturity and
\(\hat{\pi}(p) \in \erre^N\) a the vector of option prices, with the
same strike prices, depending on a parameter \(p\), the vector of
absolute relative pricing errors is
\(e=(e_1,\ldots,e_N) \in \erre^N\),
\[
  e_j := \abs[\bigg]{ \frac{\hat{\pi}_j}{\pi_j} - 1 }
  \qquad \forall j=1,\ldots,N.
\]
The loss function used to calibrate the Hermite estimator is the
\(\ell^2\) norm (i.e. the Euclidean norm) to optimize with respect to
\((\alpha)_k\) for given \(a,b\), and the \(\ell^1\) norm to optimize
with respect to \(a,b\). 
Further aspects related to the construction of Hermite approximations
will be discussed in \S\ref{sec:Had}.


\section{A variance-gamma estimator}
\label{sec:VG}
The variance-gamma estimator of option prices is constructed assuming
that logarithmic returns follow a variance-gamma process with drift
and calibratig its parameters to a set of observed data.

Let us first recall the basic definitions related to variance-gamma
processes. The gamma measure on the Borel \(\sigma\)-algebra of
\(\erre_+\) with parameters \(c, \alpha \in \erre_+^\times\), denoted
by \(\lambda_{c,\alpha}\), is defined by
\[
  \lambda_{c,\alpha}(A) := \frac{\alpha^c}{\Gamma(c)}
  \int_A x^{c-1} e^{-\alpha x}\,dx,
\]
where \(\Gamma(\cdot)\) denotes the
\href{https://dlmf.nist.gov/5}{gamma function}.
Let \(c, \alpha \in \erre_+^\times\).  Since gamma measures are
infinitely divisible, there exists a positive increasing L\'evy
process \(\Gamma\) with \(\Gamma_0=0\) such that, for every
\(t \in \erre_+\), the law of \(\Gamma_t\) is \(\lambda_{ct,\alpha}\).
We can now define variance-gamma processes as Wiener processes with
drift subordinated to gamma processes. Let \(W\) be a standard Wiener
process independent of \(\Gamma\), and \(\theta \in \erre\),
\(\sigma \in \erre_+^\times\).
The stochastic process \(Y\) defined by
\[
  Y_t = \theta \Gamma_t + \sigma W_{\Gamma_t}
\]
is a L\'evy process called (asymmetric) variance-gamma process, that
was introduced, at least in financial modeling, in \cite{MCC}, under
the extra assumption that \(c=\alpha\). We shall also use this
simplification in our numerical work, but we continue the brief
discussion in this section without assuming it.
For every \(t \in \erre_+^\times\), the law of the random variable
\(Y_t\) is absolutely continuous with a density that admits an
explicit representation in terms of the
\href{https://dlmf.nist.gov/10.25}{modified Bessel function of second
  kind} (this observation is also due to \cite{MCC}), that can be
written as
\[
  K_\nu(x) = \frac12 (x/2)^\nu \int_0^\infty t^{-\nu-1}
  \exp \bigl( -t - x^2/(4t) \bigr) \,dt
  \qquad 
\]
for \(x \in \erre\).
\begin{lemma}
  \label{lm:VGpdf}
  Let \(t \in \erre_+^\times\). The law of the random variable
  \(\theta \Gamma_t + \sigma W_{\Gamma_t}\) is absolutely continuous
  with density
  \begin{align*}
    f_{Y_t}(x)
    &= \int_0^\infty \frac{1}{\sigma \sqrt{s}} \phi \biggl(
    \frac{x-\theta s}{\sigma \sqrt{s}} \biggr)
      \,d\lambda_{ct,\alpha}(s)\\
    &= \frac{2}{\sigma\sqrt{2\pi}} \frac{\alpha^{ct}}{\Gamma(ct)}
    \exp\bigl( x\theta/\sigma^2 \bigr)
    \biggl( \frac{x^2}{\theta^2+2\alpha\sigma^2} \biggr)^{ct/2-1/4}
    K_{ct-1/2} \biggl( \abs{x} \frac{\sqrt{\theta^2 + 2\alpha\sigma^2}}%
                                     {\sigma^2} \biggr).
  \end{align*}
\end{lemma}
\noindent A proof can be inferred from the one in \cite{MCC}.

In order to guarantee that the density (at a fixed time) of a
variance-gamma process admits an expansion in Hermite functions, we
need sufficient conditions for such density to be square integrable.
\begin{prop}
  Let \(t \in \erre_+^\times\) be such that
  \[
    ct > \frac14.
  \]
  Then the density of \(Y_t\) belongs to \(L^2\).
\end{prop}
\noindent The proof is a consequence of Minkowski's inequality applied
to the first representation of the density of \(Y_t\) in Lemma
\ref{lm:VGpdf}. Details, as well as a more general result, will appear
in \cite{cm:VG}. It is not clear (to us) whether the condition is only
sufficient or also necessary.

\medskip

We can now discuss the modeling of financial returns by variance-gamma
processes (with an additional drift term) and the corresponding pricing
of European put options. We shall assume from now on that
\[
  \theta + \frac{\sigma^2}{2} < \alpha.
\]
Setting
\[
  \eta = c \log \left( 1 - \frac{\theta+\sigma^2/2}{\alpha} \right),
\]
let us assume that the process \(S\) defined by
\[
  S_t = S_0 \exp\bigl( -qt + Y_t + \eta t \bigr)
\]
is the discounted price process of an asset with constant dividend
rate \(q\) with respect to a pricing measure \(\bQ\). A simple
computation based on the tower property of conditional expectation and
the moment generating function of gamma measures shows that \(S\) is a
\(\bQ\)-martingale. 

Assuming that the short rate is nonrandom and equal to a constant
\(r\), the price at time zero of a European put option with strike
price \(K\) and expiration time \(t\) on the asset with discounted
price process \(S\) is given by
\[
e^{-rt} \Q\Bigl( K - S_0 \exp \bigl(
    (r-q)t + \eta t + Y_t \bigr)\Bigr)^+.
\]
A more explicit expression can be obtained introducing the functions
\(\tilde{q},\tilde{\sigma}\colon \erre_+ \to \erre\) defined by
\[
  \tilde{q}(x) := - (r-q+\eta)t - \bigl( \theta + \sigma^2/2 \bigr) x,
  \qquad \tilde{\sigma}(x) := \sigma \sqrt{x}.
\]
Then, using the notation of \S\ref{sec:prel}, the price of the option
is
\[
  e^{-rt} \Q \mathsf{BS}%
    \bigl(S_0,K,0,\tilde{q}(\Gamma_t),\tilde{\sigma}(\Gamma_t),1\bigr).
\]
The computation of option prices is thus reduced to the computation of
the expectation of a function of the gamma-distributed random variable
\(\Gamma_t\), that can be written as an integral. In fact, if
\(h\colon \erre_+ \to \erre\) is the function defined by
\[
h(x) := e^{-rt} \mathsf{BS}(S_0,K,0,\tilde{q}(x),\tilde{\sigma}(x),1),
\]
then the price of the put option is given by
\begin{equation}
  \label{eq:iVG}
  \int_0^\infty h(x) f_{\Gamma_t}(x)\,dx = \frac{\alpha^{ct}}{\Gamma(ct)}
  \int_0^\infty h(x) x^{ct-1} e^{-\alpha x}\,dx.
\end{equation}
In the empirical work of later sections we shall use this formula,
computing the integral numerically, to price options in the
variance-gamma setting. Since a term in the integrand has a
singularity at zero for \(t < 1/c\), i.e. for options near expiration,
the task is not entirely trivial. Some issues related to the numerical
approximation of this type of integrals are discussed in
\cite{cm:VG}.

Once a pricing formula for European options is available, calibration
of the parameters is straightforward, at least theoretically: if
\(\pi \in \erre^N\) is a vector of observed prices of options with a
fixed time to maturity and
\(\hat{\pi}(\theta,\sigma,\alpha) \in \erre^N\) is the vector of
option prices, with the same strike prices, implied by a
variance-gamma model with parameters \(\theta\), \(\sigma\), and
\(\alpha\), then we minimize the \(\ell^1\) norm of absolute relative
pricing errors with respect to \((\theta,\sigma,\alpha)\).  Note that,
in contrast to the Hermite estimator, there is no linearity in any one
of the parameters to calibrate. A few practical issues pertaining to
the calibration of the variance-gamma estimator are discussed in the
appendix.


\section{A Heston estimator}
\label{sec:Heston}
In analogy to the previous section, the Heston estimator of option
prices is constructed assuming that asset prices follow a so-called
Heston stochastic volatility model, the parameters of which are
calibrated to a set of observed option prices. This model is more
``expensive'' than the variance-gamma one, as it is specified by five
rather than just three parameters.

We shall briefly recall the definition of the Heston model and
establish an integrability result for the characteristic function of
logarithmic returns, that will imply the square integrability of
Heston densities.
Under a pricing measure \(\bQ\), let \(W=(W^1,W^2)\) be an
\(\erre^2\)-valued Wiener process with covariance matrix
\[
Q =
\begin{bmatrix}
  1 & \rho\\
  \rho & 1
\end{bmatrix},
\qquad \rho \in \mathopen]-1,1\mathclose[,
\]
and \(S_0,v_0,\kappa,\theta,\eta \in \erre_+^\times\) be such that the
so-called Feller condition
\[
  2\kappa \theta > \eta^2
\]
is satisfied. Then the system of stochastic differential equations
\begin{align*}
  S_t &= S_0 - \int_0^t qS_s\,ds + \int_0^t \sqrt{v_s} S_s\,dW^1_s,\\
  v_t &= v_0 + \int_0^t \kappa (\theta - v_s)\,ds
        + \int_0^t \eta \sqrt{v_s}\,dW_s^2
\end{align*}
admits a unique solution, for which the Feller condition implies that
\(\bQ(v_t > 0)=1\) for every \(t \geq 0\).  Interpreting \(S\) as the
discounted price process of an asset with constant dividend rate \(q\)
and \(v\) as its squared volatility process, this model was introduced
in \cite{Heston} and is since then called the Heston (stochastic
volatility) model. Note that the squared volatility \(v\) is a Feller
square-root process, better known as Cox-Ingersoll-Ross process in the
financial literature.
The Heston model is hence a parametric model of asset prices with five
parameters, namely \(v_0,\kappa,\theta,\eta,\rho\), that admit the
following interpretations: \(v_0\) is the initial squared volatility;
\(\theta\) is the ``equilibrium'' (long-time) level of the squared
volatility; \(\kappa\) is the rate of mean reversion of \(v\) towards
its equilibrium level \(\theta\); \(\eta\) is the volatility of
volatility; \(\rho\), as already mentioned, is just the correlation of
the two driving Wiener processes.

A convenient technique to price options on assets with price process
following a Heston model is via Fourier transform methods. In the
following we shall assume, as done before, without loss of generality,
that \(q=0\) and \(S_0=1\).  General results on affine processes (see,
e.g., \cite{DuFiScha}), or the original arguments in \cite{Heston}),
imply that there exist functions
\(A,B\colon \erre \times \erre_+ \to \bC\) such that the
characteristic function \(\psi_t\) of \(\log S_t\) can be written
as
\[
  \psi_t(\xi) := \bQ \exp\bigl( i \xi \log S_t \bigr)
  = \exp \bigl( A(\xi,t) + B(\xi,t)v_0 \bigr).
\]
Moreover, the functions \(A,B\) admit explicit expressions. Namely,
introducing the functions \(\alpha,\beta\colon \erre \to \bC\) defined
by
\[
  \widehat{\alpha}(\xi) := -\frac12 \xi(\xi+i), \qquad
  \beta(\xi) := \kappa - i \eta \rho \xi,
\]
setting \(\gamma := \eta^2/2\), and introducing the functions
\(D,G\colon \erre \to \bC\) defined by
\[
  D := \bigl( \beta^2 - 4\gamma\widehat{\alpha} \bigr)^{1/2},
  \qquad G := \frac{\beta-D}{\beta+D},
\]
one has
\begin{align*}
  A(\xi,t)
  &:= \frac{\kappa\theta}{\eta^2} \biggl( (\beta(\xi) + D(\xi))t %
    - 2\log \frac{e^{D(\xi)t} - G(\xi)}{1 - G(\xi)} \biggr),\\
  B(\xi,t)
  &:= \frac{\beta(\xi) - D(\xi)}{\eta^2} \,
    \frac{1-e^{-D(\xi)t}}{1-G(\xi) e^{-D(\xi)t}}.
\end{align*}
In the expression for \(A\) the square root and the logarithm of a
complex number \(z = \abs{z} (\cos\vartheta + i \sin\vartheta)\),
\(\vartheta \in \mathopen[-\pi,\pi\mathclose[\), are defined as
\[
  z^{1/2} := \abs{z}^{1/2}\bigl( \cos \vartheta/2 + i \sin \vartheta/2
  \bigr), \qquad
  \log z := \log \abs{z} + i \vartheta
\]
(the square root on the right-hand side of the first definition being
in the sense of real numbers).

We shall need a criterion for the density of \(\log S_t\) to belong to
\(L^2\). The following lemma is just a technical step.
\begin{lemma}
  \label{lm:air}
  Let \(\widetilde{A}\colon \erre \times \erre_+ \to \bC\) be the
  function defined by
  \[
    \widetilde{A}(\xi,t) := \frac{\kappa\theta}{\eta^2} \biggl(
    (\beta(\xi) - D(\xi))t - 2\log \frac{G(\xi)e^{-D(\xi)t} -
      1}{G(\xi) - 1} \biggr).
  \]
  Then \(\widetilde{A} = A\) \(\operatorname{mod} i\erre\).
\end{lemma}
\begin{proof}
  Writing \(Dt = -Dt + 2Dt\), since
  \(2Dt = -2 \log e^{-Dt}\) \(\operatorname{mod} i\erre\), one has
  \[
    (\beta + D)t - 2\log \frac{e^{Dt}/G-1}{1/G-1}
    = (\beta - D)t - 2\log e^{-Dt} - 2\log \frac{G-e^{Dt}}{G-1}
    \qquad \operatorname{mod} i\erre,
  \]
  where
  \[
    \log e^{-Dt} + \log \frac{G-e^{Dt}}{G-1} =
    \log \frac{Ge^{-Dt}-1}{G-1} \qquad \operatorname{mod} i\erre,
  \]
  thus establishing the claim.
\end{proof}

We can now establish an integrability property of the
characteristic function \(\psi_t\).
\begin{prop}
  The characteristic function \(\psi_t\) of \(\log S_t\) belongs to
  \(L^p\) for every \(p \in [1,+\infty]\).
\end{prop}
\begin{proof}
  By the \href{https://en.wikipedia.org/wiki/Riemann%E2%80%93Lebesgue_lemma}{Riemann-Lebesgue lemma}, \(\psi_t\) is uniformly continuous
  and null at infinity, in particular it is bounded.  Therefore it is
  enough to show that \(\psi_t\) is \(p\)-integrable outside an
  arbitrarily large compact set. In view of Lemma \ref{lm:air}, the
  functions \(A\) and \(\widetilde{A}\) differ only by an imaginary
  term, hence it suffices to show that
  \(\exp (\widetilde{A}+B) = \exp\widetilde{A} \, \exp B\) belongs to
  \(L^p\).
  Setting \(c:=\kappa\theta/\eta^2\) for convenience, one has
  \[
    \exp \bigl( \widetilde{A}(\xi,t) \bigr) = \biggl(
    \frac{G(\xi)-1}{G(\xi)e^{-D(\xi)t}-1} \biggr)^{2c} \exp\bigl(
    c(\beta(\xi) - D(\xi))t \bigr).
  \]
  Since
  \begin{align*}
    \beta^2 - 4\gamma\widehat{\alpha}
    &= (\kappa - i\eta\rho\xi)^2 +\eta^2 \xi(\xi+i)\\
    &= \kappa^2 + (1-\rho^2)\eta^2 \xi^2 + i \eta(\eta-2\kappa\rho)\xi,
  \end{align*}
  with \(\abs{\rho}<1\), the real part of
  \(\beta^2 - 4\gamma\widehat{\alpha}\) is positive. This in turn
  implies that the real part of
  \((\beta^2 - 4\gamma\widehat{\alpha})^{1/2}\) is positive. More
  precisely, setting
  \(\beta^2 - 4\gamma\widehat{\alpha} =: z =: \abs{z}e^{i\vartheta}\),
  one has
  \begin{align*}
    \abs{z}^{1/2}
    &= \bigl( \kappa^4 + 2\kappa^2(1-\rho)^2\eta^2\xi^2
    + (1-\rho^2)^2\eta^4\xi^4 + \eta^4 (\eta-2\kappa\rho)^4\xi^4
      \bigr)^{1/4},\\
    \vartheta
    &= \arctan \frac{\eta(\eta-2\kappa\rho)\xi}%
    {\kappa^2 + (1-\rho^2)\eta^2\xi^2},
  \end{align*}
  hence
  \((\beta^2 - 4\gamma\widehat{\alpha})^{1/2} = \abs{z}^{1/2}
  e^{i\vartheta/2}\).  Since
  \(\lim_{\xi\to\infty} \vartheta(\xi) = 0\), there exists
  \(\xi_0 \in \erre_+\) such that, for every \(\xi \in \erre\) with
  \(\abs{\xi}>\xi_0\),
  \[
    \operatorname{Re} \bigl( \beta(\xi) - D(\xi) \bigr) \eqsim 1 - \abs{\xi}.
  \]
  Let us now show that the function \(G-1\) is bounded. In fact,
  \[
    \abs{G(\xi)-1} = 2 \frac{\abs{D(\xi)}}{\abs{\beta(\xi) +
        D(\xi)}},
  \]
  where \(\abs{D(\xi)}^2 \eqsim 1 + \xi^2\) and
  \[
    \beta(\xi) + D(\xi) = \kappa - i\eta\rho\xi
    + \abs{z}^{1/2} ( \cos \vartheta/2 + i \sin
    \vartheta/2 ),
  \]
  where, entirely similarly as above,
  \(\abs{z}^{1/2}(\xi) \eqsim 1 + \abs{\xi}\) for \(\abs{\xi}\)
  sufficiently large, and
  \(\lim_{\xi \to \infty} \vartheta(\xi) = 0\), from which it
  immediately follows that
  \(\abs{\beta(\xi) + D(\xi)} \eqsim 1 + \abs{\xi}\) for \(\xi\)
  outside a compact set.  Therefore the function \(G-1\), as well as
  \(G\) itself, are bounded outside a compact set.
  Since the real part of \(D(\xi)\), as seen above, is proportional to
  \(1+\abs{\xi}\) for \(\abs{\xi}\) large, this also implies that
  \[
    \lim_{\xi \to \infty} \abs[\big]{G(\xi)e^{-D(\xi)t}} = 0,
  \]
  thus also that
  \[
    \lim_{\xi \to \infty} \abs[\big]{G(\xi)e^{-D(\xi)t} - 1} = 1,
  \]
  from which it follows that the function \((G-1)/(Ge^{-Dt}-1)\) is
  bounded outside a compact set.
  Therefore
  \(\operatorname{Re} \exp A(\xi,t) \eqsim \exp (c(1-\abs{\xi}))\) for
  every \(\xi\) outside a compact set.
  Concerning \(B\), note that, thanks to the previous estimates,
  \[
    \lim_{\xi \to \infty} \frac{1-e^{-D(\xi)t}}{1-G(\xi)e^{-D(\xi)t}}
    = 1,
  \]
  hence, recalling that
  \(\operatorname{Re} (\beta(\xi) - D(\xi)) \eqsim 1 - \abs{\xi}\) for
  \(\abs{\xi}\) large, it follows that
  \(\operatorname{Re} B(\xi,t) \eqsim 1 - \abs{\xi}\) outside
  a compact set.
  We have thus shown that, for \(\xi\) outside a compact set, there
  exists a strictly positive constant \(C\) such that
  \(\psi_t(\xi) \eqsim \exp \bigl( -C \abs{\xi} \bigr)\), which
  implies that \(\psi_t \in L^p\) for every \(p \in [1,+\infty]\).
\end{proof}

\begin{coroll}
  \label{cor:HL2}
  The density of the random variable \(\log S_t\) belongs to \(L^2\).
\end{coroll}
\begin{proof}
  Immediate recalling that the Fourier transform is an (isometric)
  isomorphism of \(L^2\).
\end{proof}

As the characteristic function of \(\log S_t\) is known, prices of put
options can be efficiently computed by an approach, by now classical,
apparently first proposed in \cite{CaMa:FFT}. In particular, denoting
the logarithm of the strike price by \(k\) and the law of the random
variable \(\log S_t\) by \(\mu\), the price of a put option with
strike price \(e^k\) can be written as
\[
p(k) := \int_{-\infty}^k (e^k-e^x) \,d\mu(x),
\]
for which it is easily seen that, for any \(\alpha \leq -1\), the
function \(p_\alpha\) defined by
\[
  p_\alpha(k) := e^{\alpha k} p(k),
\]
belongs to \(L^1\). Therefore, introducing the Fourier transform of
\(p_\alpha\) defined by
\[
\cF p(\alpha)(\xi) := \int_\erre e^{-2i\pi\xi k} p_\alpha(k)\,dk,
\]
if \(\cF p_\alpha \in L^1\), then, by inverse Fourier transform, one
has, using well-known properties of Fourier tranforms of real-valued
functions,
\begin{align*}
  p(k) = e^{-\alpha k} p_\alpha(k) %
  &= e^{-\alpha k} \int_\erre e^{2i\pi\xi k} \cF p_\alpha(\xi)\,d\xi\\
  &= 2 e^{-\alpha k} \int_0^\infty e^{2i\pi\xi k} \cF p_\alpha(\xi)\,d\xi\\
  &= 2 e^{-\alpha k} \int_0^\infty \operatorname{Re} e^{2i\pi\xi k}
    \cF p_\alpha(\xi)\,d\xi.
\end{align*}
A simple computation based on Tonelli's theorem yields
\[
  \cF p_\alpha(\xi) = \frac{\cF\mu\bigl(\xi + i(\alpha+1)/(2\pi)\bigr)}%
  {(\alpha+1-i2\pi\xi) (\alpha-i2\pi\xi)},
\]
therefore, setting
\[
  \psi(\xi,\alpha) := \operatorname{Re}%
  \frac{\cF\mu\bigl(\xi + i(\alpha+1)/(2\pi)\bigr)}%
  {(\alpha+1-i2\pi\xi) (\alpha-i2\pi\xi)}
\]
for convenience of notation, one has
\[
p(k) = 2 e^{-\alpha k} \int_0^\infty \psi(\xi,\alpha)\,d\xi.
\]
Even though the right-hand side does not depend on \(\alpha\), the
choice of \(\alpha\) influences the accuracy of numerical
integration. We shall adopt the methods of \cite{LoKa07}, selecting
\(\alpha\) in an automated payoff-dependent way (cf. \cite[p.~13,
eq.~(60)]{LoKa07}), reducing the integral over \(\erre_+\) to an
integral over a finite domain thanks to precise asymptotic relations
for the function \(\psi\), for which we refer to
\cite[Proposition~2.2]{LoKa07}.

Calibration of Heston models to observed option prices is done,
\textit{mutatis mutandis}, as already described at the end of
\S\ref{sec:VG}. Some related numerical issues are discussed in the
appendix.


\section{Hermite approximation of densities}
\label{sec:Had}
Before looking at the empirical performance of variance-gamma and
Heston estimators compared with some Hermite estimators with the same
number of parameters, it seems interesting, as a preliminary step, to
examine whether variance-gamma and Heston densities can be
approximated by truncated expansions in Hermite functions, how good
the approximations are in \(L^2\) norm, and whether Hermite
approximations can estimate option prices implied by the parametric
densities within reasonable error bounds. While numerical experiments
are deferred to \S\S\ref{ssec:VGtoH}-\ref{ssec:HestoH}, in
\S\ref{ssec:Hlr} we discuss some issues regarding expansions in series
of Hermite functions of arbitrary densities of logarithmic returns,
showing that optimizing the scale and location parameters is feasible
and that some classes of constrained Hermite approximations can be
computed explicitly from unconstrained ones. In
\S\S\ref{ssec:VGd}-\ref{ssec:Hesd} we deal with the computation of the
coefficients \((\alpha_k)\) of Hermite approximations to
variance-gamma and Heston densities, respectively, showing that in the
former case the \((\alpha_k)\) admit closed-form expressions in terms
of integrals with respect to gamma measures, while in the latter case
they can be written as the scalar product in \(L^2\) of Hermite
functions and a simple modification of the Fourier transform of the
density.

\subsection{Densities of logarithmic returns}
\label{ssec:Hlr}
According to the usual conventions, let \(S_t\) denote the discounted
price of an asset at time \(t\). We shall assume, for the purposes of
this subsection, that \(t\) is fixed, \(q=0\), \(S_0=1\),
\(\Q(S_t>0)=1\), and that the law of \(\log S_t\) with respect to
\(\bQ\) admits a density \(f\) belonging to \(L^2\).

Since \(f\) belongs to \(L^2\), given an orthonormal basis \((h_n)\)
of \(L^2\), one can always write
\(f = \sum_{n \in \bN} \alpha_n h_n\), with
\(\alpha_n := \ip{f}{h_n}\) and the identity to be understood in the
\(L^2\) sense. If \((h_n)\) are (properly scaled) Hermite functions, a
truncation of the series representing \(f\) provides a
finite-dimensional approximation that, however, turns out to be too
crude for the purposes of estimating option prices. It is instead much
preferable to formally set \(\log S_t = a X + b\), for some constants
\(a, b\) (the choice of which will be discussed next) and a random
variable \(X\), and to approximate the density of \(X\), rather than
the density of \(\log S_t\).
Since the density of \(X\) is easily seen to be
\(x \mapsto af(ax+b)\), setting \(f_{a,b} \colon x \mapsto f(ax+b)\)
and
\[
  \alpha_k := \alpha_k(a,b)
  := \frac{1}{{\norm{h_k}}_{L^2}^2} \ip{f_{a,b}}{h_k}
  = \frac{1}{k! \sqrt{\pi}} \ip{f_{a,b}}{h_k}
  \qquad \forall k \in \bN,
\]
so that
\[
f_{a,b} = \sum_{k=0}^\infty \alpha_k h_k
\]
as an identity in \(L^2\), one has
\[
  a \int_\erre \abs[\bigg]{f(ax+b) - \sum_{k=0}^\infty \alpha_k
  h_k(x)}^2\,dx
  = \int_\erre \abs[\bigg]{f(x) - \sum_{k=0}^\infty \alpha_k
    h_k\left( \frac{x-b}{a} \right)}^2\,dx,
\]
from which it is immediate that, for every \(n \in \bN\), one can
approximate the density of \(\log S_t\), in the \(L^2\) sense, by the
function \(f_n\) defined by
\[
f_n(x) := \sum_{k=0}^n \alpha_k h_k\left( \frac{x-b}{a} \right),
\]
that is the projection of \(f\) on the finite-dimensional subspace of
\(L^2\) spanned by the functions \((h_k((\cdot-b)/a))_{k=0,\ldots,n}\).

Let us consider the issue of choosing the constants \(a\) and \(b\).
As already mentioned (cf. also \cite{cm:Herm}), if one interprets the Hermite
approximation of the density of \(X\) as a perturbation of the
Black-Scholes model for asset prices, in the sense that the Hermite
approximation of order zero coincides with the Black-Scholes model,
for which \(\log S_t = \varsigma Z - \varsigma^2/2\) in law, with
\(Z\) standard Gaussian random variable and
\(\varsigma \in \erre_+^\times\), then the constant \(a\) and \(b\)
would have to be chosen as
\[
b = -a^2/2, \qquad b = \Q \log S_t,
\]
or, alternatively, as
\[
  b = \Q \log S_t, \qquad
  a = \operatorname{Std}_\bQ \log S_t
  := \bigl( \Q(\log S_t - b)^2 \bigr)^{1/2}.
\]
Note that these two choices of the constants \(a\) and \(b\) are
equivalent if \(\log S_t = \varsigma Z - \varsigma^2/2\), but are in
general \emph{not} equivalent. Moreover, the martingale property of
\(S\) and Jensen's inequality imply that \(\Q \log S_t \leq 0\), so
there is no issue with the former choice of constants. On the other
hand, in order for the standard deviation of \(\log S_t\) to be well
defined, it is necessary to assume that \(\Q(\log S_t)^2 < \infty\)
or, equivalently, \(\int_\erre x^2 f(x)\,dx<\infty\). This condition
is automatically satisfied if, for instance, \(f\) is symmetric, as in
this case
\[
  \int_\erre x^2 f(x)\,dx = 2\int_{\erre_+} x^2 f(x)\,dx
  \leq 2\int_{\erre_+} e^x f(x)\,dx \leq 2\Q S_t = 2.
\]
It is also easy to check that
\[
\int x^2 f_n(x)\,dx < \infty \qquad \forall n \in \bN.
\]
A further natural step is to select the constants \(a\) and \(b\) by an
optimization procedure, i.e., setting
\[
f_n^{s,m} := \sum_{k=0}^n \alpha_k(s,m) h_k \bigl( (\cdot - m)/s \bigr)
\]
for convenience of notation, and
\(J(s,m) := \norm[\big]{f - f_n^{s,m}}_{L^2}^2\),
to consider the problem
\[
  \inf_{(s,m) \in \erre_+^\times \times \erre} J(s,m)
\]
and, assuming that a minimizer \((m_\ast,s_\ast)\) exist, to set
\(a := s_\ast\) and \(b := m_\ast\). For the minimization of \(J\) it
is useful to show that it is smooth, which we do next.
We start with a simple bound that is going to be used often.
\begin{lemma}
  \label{lm:pobo}
  Let \(A_1 \subset \erre_+^\times\), \(A_2, A_3 \subset \erre\) be
  compact sets, \(A:=A_1 \times A_2 \times A_3\), and
  \((P_i)_{i \in I}\) be an arbitrary collection of polynomials of
  degree bounded by \(n \in \bN\), such that their coefficients belong
  to \(A_3\).  There exists a function \(g\colon \erre \to \erre_+\),
  belonging to \(L^p\) for every \(p \in [1,\infty]\), such that, for
  every \(x \in \erre\), \(i \in I\), \(a \in A_1\), and
  \(b \in A_2\), one has
  \[
    \abs[\bigg]{P_i(x) \phi\biggl( \frac{x-a}{b} \biggr)}
    \lesssim_{n,A} g(x).
  \]
\end{lemma}
\begin{proof}
  Let \(i \in I\). The polynomial \(P_i\) is the sum of monomials
  \(c_k x^k\), where \(k \in \bN\), \(k \leq n\), hence
  \(\abs{x}^k \leq 1 + \abs{x}^n\) for all \(x \in \erre\), and there
  exists \(c \in \erre_+^\times\) such that \(\abs{c_k} \leq c\). Then
  \(\abs{P_i(x)} \leq cn\bigl( 1 + \abs{x}^n \bigr)\).
  Moreover, setting \(a_\ast := \max_{a \in A_1} \abs{a}\) and
  \(b_\ast := \max_{b \in A_2} \abs{a}\), one has, for every
  \(x \in \erre\),
  \[
    -\frac12 \frac{(x-b)^2}{a^2} \leq -\frac12
    \frac{b^2 - 2bx + x^2}{a_\ast^2}
  \]
  which implies
  \[
    \exp\biggl( -\frac12 \frac{(x-b)^2}{a^2} \biggr)
    \leq \exp\bigl( -b^2/(2a_\ast^2)\bigr)
    \exp\bigl( bx/a_\ast^2 \bigr)
    \exp\bigl( -x^2/(2a_\ast^2)\bigr),
  \]
  where the first term on the right hand side is bounded by a constant
  depending only on \(A\), and the second is bounded by
  \(\exp(b_\ast\abs{x}/a_\ast^2)\). Therefore, defining the function \(g\)
  by
  \[
    g(x) := \exp\bigl( -x^2/(2a_\ast^2) + b_\ast\abs{x}/a_\ast^2 \bigr) (1 +
    \abs{x}^n),
  \]
  that is easily seen to belong to \(L^p\) for every
  \(p \in [1,\infty]\), it follows that \(\abs{f_n} \lesssim g\)
  pointwise with an implicit constant depending only on \(n\) and
  \(A\), as claimed.
\end{proof}

\begin{coroll}
  \label{cor:pobo}
  Let \(s_0,s_1,m_0 \in \erre_+^\times\) with \(s_0 < s_1\). There
  exists a function \(g\colon \erre \to \erre_+\), belonging to
  \(L^p\) for every \(p \in [1,\infty]\), such that, for every
  \(x \in \erre\), \(s \in [s_0,s_1]\), and \(m \in [-m_0,m_0]\), one
  has
  \[
    \abs{f_n^{s,m}(x)} \lesssim g(x) \qquad \forall x \in \erre,
  \]
  where the implicit constant depends only on \(n\), \(s_0\), \(m_0\),
  and \({\norm{f}}_{L^2}\).
\end{coroll}
\begin{proof}
  In the identity
  \[
    f_n^{s.m}(x) = \sqrt{2\pi} \sum_{k=0}^n \alpha_k(s,m)
    H_k\biggl( \frac{x-m}{s} \biggr) \phi\biggl( \frac{x-m}{s} \biggr)
  \]
  one has, by the Cauchy-Schwarz inequality, that, for every
  \(k \leq n\),
  \[
    \abs{\alpha_k(s,m)} = \abs[\big]{\ip{f(s \cdot + m)}{h_k}}
    \leq \norm[\big]{f(s\cdot + m))}_{L^2} \norm[\big]{h_k}_{L^2}
    = \frac{\sqrt{k!\sqrt{\pi}}}{\sqrt{s}} \norm[\big]{f}_{L^2},
  \]
  i.e. \(\abs{\alpha_k}\) is bounded by a constand depending on \(n\),
  \(s_0\), and \(\norm{f}_{L^2}\). Moreover, for every \(k \leq n\),
  \(H_k((x-m)/s\) is a polynomial of degree \(k\) and coefficients of
  the type \(m^p/s^q\), with \(p,q \in \bN\). Since
  \(m^p/s^q \leq m_0^p/s_0^q\), it immediately follows that
  \[
    \sum_{k=0}^n \alpha_k(s,m) H_k\biggl( \frac{x-m}{s} \biggr)
  \]
  is a polynomial in \(x\) with coefficients that are uniformly
  bounded by a constant depending on \(n\), \(s_0\), \(m_0\), and
  \({\norm{f}}_{L^2}\). The claim then follows by Lemma \ref{lm:pobo}.
\end{proof}

\begin{lemma}
  \label{lm:hkdiff}
  For any \(k \in \bN\), the function
  \begin{align*}
    \erre_+^\times \times \erre &\longrightarrow L^2\\
    (s,m) &\longmapsto h_k\biggl( \frac{\cdot - m}{s} \biggr)
  \end{align*}
  is of class \(C^\infty\).
\end{lemma}
\begin{proof}
  The function \(s \mapsto (x-m)/s\) is of class \(C^\infty\) for
  every \(x,m \in \erre\) and the function \(m \mapsto (x-m)/s\) is of
  class \(C^\infty\) for every \(x \in \erre\) and every
  \(s \in \erre_+^\times\). Since \(H_k\) and \(\phi\) are both of
  class \(C^\infty\), it follows by composition that
  \((s,m) \mapsto h_k((s-m)/s)\) are of class \(C^\infty\) for every
  \(x \in \erre\).  Therefore, if \((s_n,m_m)\) is a sequence in
  \(\erre_+^\times \times \erre\) converging to \((s,m)\), this
  implies, thanks to Corollary \ref{cor:pobo} and the dominated convergence
  theorem, that
  \[
    \lim_{n \to \infty} \int_\erre
    \abs[\big]{h_k((x-m_n)/s_n) - h_k((x-m)/s)}^2\,dx = 0,
  \]
  i.e. that the function \((s,m) \mapsto h_k((\cdot-m)/s)\) is
  continuous.  Moreover, the function \(s \mapsto h_k((x-m)/s)\) is
  differentiable for every \(x \in \erre\) with derivative
  \[
    s \mapsto -\frac{1}{s^2} (x-m) h'_k\biggl( \frac{x-m}{s} \biggr),
  \]
  that is
  \[
    \lim_{h \to 0} \frac1h \bigl( h_k( (x-m)/(s+h) ) - h_k( (x-m)/s ) \bigr)
    + \frac{1}{s^2} (x-m) h'_k( (x-m)/s ) = 0
  \]
  for every \(x \in \erre\). The fundamental theorem of calculus yields
  \[
    h_k( (x-m)/(s+h) ) - h_k( (x-m)/s ) = \int_0^h \frac{1}{(s+t)^2}
    (x-m) h'_k((x-m)/(s+t))\,dt,
  \]
  where, by Lemma \ref{lm:pobo}, for \(h\) sufficiently small the
  absolute value of the integrand on the right-hand side is bounded
  uniformly with respect to \(t\) by an \(L^2\) function \(g\) (of
  \(x\)), so that
  \[
    \frac1h \abs[\big]{h_k( (x-m)/(s+h) ) - h_k( (x-m)/s )}
    \leq \frac1h \int_0^h g(x)\,dt = g(x).
  \]
  The dominated convergence theorem then allows to conclude that
  \[
    \lim_{h \to 0} \int_\erre \abs[\bigg]{
      \frac1h \bigl( h_k( (x-m)/(s+h) ) - h_k( (x-m)/s ) \bigr)
    + \frac{1}{s^2} (x-m) h'_k( (x-m)/s )}^2\,dx = 0,
  \]
  thus proving that \(s \mapsto h_k((\cdot-m)/s)\) is differentiable
  with derivative
  \[
    s \mapsto - \frac{1}{s^2} (\cdot-m) h'_k((\cdot-m)/s).
  \]
  Entirely similar (in fact slightly simpler) arguments show that
  \(m \mapsto h_k((\cdot-m)/s)\) is differentiable with derivative
  \[
    m \mapsto - \frac{1}{s} h_k'((\cdot-m)/s).
  \]
  A further application of Lemma \ref{lm:pobo} shows that the partial
  derivatives of the function \((s,m) \mapsto h_k(\cdot-m)/s)\) are
  continuous, hence that the function is Fr\'echet differentiable
  and of class \(C^1\).
  Moreover, the function \((s,m) \mapsto h_k((x-m)/s)\) has continuous
  partial derivatives of every order for every \(x \in \erre\) that
  can be written as a polynomial (with coefficients given by
  continuous functions of \(m\) and \(1/s\)) multiplied by the density
  function of Gaussian random variable with mean \(m\) and standard
  deviation \(s\), from which it follows, by repeated applications of
  Lemma \ref{lm:pobo} and the dominated convergence theorem, that
  \((s,m) \mapsto h_k((\cdot-m)/s)\colon \erre^2 \to L^2\) admits
  continuous partial derivatives of every order, hence that it is of
  class \(C^\infty\).
\end{proof}
\begin{rmk}
  Continuity of \((s,m) \mapsto h_k((\cdot-m)/s)\) can be proved in at
  least two other ways: in the first way as a consequence of the more
  general result according to which translations and dilations are
  strongly continuous linear operators on \(L^p\) for every
  \(p \in \mathopen[1,\infty\mathclose[\). The
  second way proceeds noting that
  \[
    \norm[\big]{h_k(\cdot-m)/s)}_{L^2} = \sqrt{s} \norm[\big]{h_k}_{L^2}
  \]
  immediately implies that
  \((s,m) \mapsto {\norm{h_k((\cdot-m)/s}}_{L^2}\) is continuous and
  bounded on every bounded subset of \(\erre_+^\times \times
  \erre\). The
  \href{https://doi.org/10.1090/S0002-9939-1983-0699419-3}{Brezis-Lieb
    lemma} (see \cite{BreLieb}) then easily yields that
  \((s,m) \mapsto h_k((\cdot-m)/s)\) is continuous from
  \(\erre_+^\times \times \erre\) to \(L^2\).
\end{rmk}

\begin{prop}
  For any \(n \in \bN\), the function
  \begin{align*}
    \erre_+^\times \times \erre &\longrightarrow L^2\\
    (s,m) &\longmapsto f_n^{s,m} =
            \sum_{k=0}^n \alpha_k(s,m) h_k\biggl( \frac{\cdot - m}{s} \biggr)
  \end{align*}
  is of class \(C^\infty\).
\end{prop}
\begin{proof}
  For any \(k \in \bN\),
  \[
    \alpha_k(s,m) = \ip[\big]{f(s\cdot + m)}{h_k}
    = \frac{1}{s} \ip[\big]{f}{h_k((\cdot-m)/s)},
  \]
  where, by Lemma \ref{lm:hkdiff},
  \((s,m) \mapsto h_k((\cdot-m)/s) \in C^\infty(\erre^2;L^2)\). Since
  the map \(\ip{f}{\cdot} \colon L^2 \to \erre\) is linear and
  continuous, hence of class \(C^\infty\), it follows immediately, by
  composition, that
  \((s,m) \mapsto \alpha_k(s,m) \in C^\infty(\erre^2)\). Therefore,
  since the product \(\erre \times L^2 \to L^2\) is bilinear and
  continuous, hence smooth, one has
  \[
    (s,m) \mapsto \alpha_k(s,m) h_k((\cdot-m)/k) \in C^\infty(\erre^2;L^2)
  \]
  for every \(k \in \bN\), from which the claim follows immediately.  
\end{proof}

\begin{prop}
  The function \(J \colon \erre_+^\times \times \erre \to \erre\) is
  of class \(C^\infty\).
\end{prop}
\begin{proof}
  The function \(J\) is the composition of
  \((s,m) \mapsto f_n^{s,m}-f \in C^\infty(\erre^2;L^2)\) and
  \({\norm{\cdot}}^2_{L^2} \in C^\infty(L^2)\), hence
  \(J \in C^\infty(\erre^2)\).
\end{proof}

Let us compute the derivatives of first and second order of \(J\). As
a first step, note that, for any \(g \in L^2\),
\(D{\norm{g}}^2_{L^2} = v \mapsto 2\ip{g}{v}\), hence, by the chain
rule for (Fr\'echet derivatives), \(DJ = 2\ip{f-f_n}{Df_n}\).  For the
purposes of this computation only, let us denote the function
\((s,m) \to h_k((\cdot-m)/s)\colon \erre^2 \to L^2\) also by \(h_k\).
Denoting the canonical basis of \(\erre^2\) by \((e_1,e_2)\), one has,
as in the proof of Lemma \ref{lm:hkdiff},
\begin{align*}
  [Dh_k]e_1
  &= (s,m) \mapsto \partial_1 h_k((\cdot-m)/s)
    = -\frac{1}{s^2} (\cdot - m) h'_k((\cdot-m)/s)\\
  [Dh_k]e_2
  &= (s,m) \mapsto \partial_2 h_k((\cdot-m)/s)
    = -\frac{1}{s} h'_k((\cdot-m)/s),
\end{align*}
hence \(Dh_k\colon \erre_+^\times \times \erre \to \cL(\erre^2;L^2)\)
is identified (by linearity) by the above expressions for
\([Dh_k]e_1\) and \([Dh_k]e_2\). Similarly,
\begin{align*}
  \partial_1 \alpha_k\colon (s,m)
  &\mapsto - \frac{1}{k!\sqrt{\pi}} \frac{1}{s^2}
    \ip[\big]{f}{h_k((\cdot-m)/s)}
    - \frac{1}{k!\sqrt{\pi}} \frac{1}{s^3}
    \ip[\big]{f}{(\cdot-m) h'_k((\cdot-m)/s)},\\
  \partial_2 \alpha_k\colon (s,m)
  &\mapsto - \frac{1}{k!\sqrt{\pi}} \frac{1}{s^2}
    \ip[\big]{f}{h'_k((\cdot-m)/s)}
\end{align*}
hence
\(D\alpha_k\colon \erre^2 \to \cL(\erre^2;\erre) \simeq \erre^2\) is
identified, thanks to linearity, by
\([D\alpha_k]e_1 = \partial_1\alpha_k\) and
\([D\alpha_k]e_2 = \partial_2\alpha_k\). Then
\(D(\alpha_k h_k) = h_k D\alpha_k + \alpha_k Dh_k\), where \(h_k\) is
interpreted as a function \(\erre^2 \to \cL(\erre;L^2)\), hence
\[
  Df_n = \sum_{k=0}^n h_k D\alpha_k + \alpha_k Dh_k
  \colon \erre_+^\times \times \erre \longrightarrow \cL(\erre^2;L^2)
\]
and \(DJ\) is identified by
\[
[DJ]e_i = 2\ip[\big]{f-f_n}{[Df_n]e_i}, \qquad i=1,2.
\]
In order to compute \(D^2J\), recall that
\((g_1,g_2) \mapsto \ip{g_1}{g_2}\colon L^2 \times L^2 \to \erre\) is, as
every continuous bilinear map, continuously differentiable, with
Fr\'echet derivative
\(D\ip{g_1}{g_2}\colon (h,k) \mapsto \ip{g_1}{k} +
\ip{g_2}{h}\). Then, by the chain rule,
\[
  D^2J = D\ip{f-f_n}{Df_n}\colon \erre^2 \to \cL(\erre^2;\cL(\erre^2;L^2))
  \simeq \cL_2(\erre^2;L^2)
\]
is the function with values in the space of symmetric bilinear forms
on \(\erre^2\) with values in \(L^2\) defined, for every
\(u,v \in \erre^2\), by
\[
  (u,v) \longmapsto -\ip[\big]{[Df_n]u}{[Df_n]v} +
  \ip[\big]{f-f_n}{[D^2f_n](u,v)}.
\]
An explicit expression for \([D^2f_n](u,v)\) can be obtained as
follows:
\[
  [D^2(\alpha_k h_k)](u,v) = D\alpha_k(u) Dh_k(v) + D\alpha_k(v) Dh_k(u)
  + D^2\alpha_k(u,v) h_k + \alpha_k D^2h_k(u,v),
\]
where, considering \(\erre^2\) with its canonical basis,
\(D^2\alpha_k\) and \(D^2h_k\) are represented by their Hessian matrices
(with real and \(L^2\)-valued entries, respectively).

Even though the objective function \(J\) is (infinitely many times)
differentiable with respect to its arguments, with explicit
expressions for its derivatives, it does not seem simple to determine
analytically whether minimizers exist and, if they do, to identify
global minima.
In practice, one relies on numerical minimization algorithms using as
initial values for \(s\) and \(m\) the values of \(a\) and \(b\)
discussed above. This procedure can be interpreted as locally
optimizing the constants in a Hermite perturbation of a starting
Gaussian (i.e. Black-Scholes) approximation of logarithmic
returns. Note that the availability of easily implementable
expressions for the gradient of the objective function, that follows
from the above results, can speed up numerical minimization algorithms
considerably. Since second derivatives are easily implementable as
well, it is also possible to check numerically in an efficient way
whether critical points for the gradient are (local) minimum points.

\medskip

We are now going to discuss a class of Hermite approximations subjects
to constraints. As motivation, note that the function \(f_n\) is
not, in general, the density of a random variable, as it may assume
negative values and may not have unit integral. While enforcing a
positivity constraint on Hermite approximations such as \(f_n\) seems
unfeasible, it is relatively simple to determine, for fixed \(a\) and
\(b\), the best \(L^2\) approximation of \(f\) of the type
\[
  \widetilde{f}_n(x) = \sum_{k=0}^n \beta_k h_k\left( \frac{x-b}{a} \right),
\]
with \((\beta_k)\) constants, such that
\[
\int_\erre \widetilde{f}_n(x)\,dx = 1.
\]
In geometric terms, the problem amounts to finding the projection of
\(f\) on the finite-dimensional closed convex subset of \(L^2\) formed
by functions such as \(\widetilde{f}_n\). This projection can be
characterized explicitly and is a consequence of the next more general
fact.

Let \(L \in \operatorname{Hom}(\erre^{n+1};\erre^m)\) be surjective
(in particular, \(m \leq n+1\)), \(v \in \erre^m\), and
\(H := \{\beta \in \erre^{n+1}:\, L\beta = v\}\). The matrix
representing the homomorphism \(L\) with respect to the canonical
bases of \(\erre^{n+1}\) and \(\erre^m\) will be denoted by \(L\) as
well.
\begin{lemma}
  Let \(\alpha \in \erre^{n+1}\). The problem
  \[
    \inf_{\beta \in H} \norm[\big]{\beta - \alpha}_{\erre^{n+1}}
  \]
  admits a solution and the infimum is attained by the vector
  \[
    \beta_\ast := \alpha - L^\top \bigl( LL^\top \bigr)^{-1} (L\alpha - v).
  \]
\end{lemma}
\begin{proof}
  Consider the Lagrangian
  \[
    F(\beta;\lambda) := \frac12
    \norm[\big]{\beta-\alpha}^2_{\erre^{n+1}} +
    \ip[\big]{\lambda}{L\beta - v}_{\erre^m}
  \]
  where \(\lambda\) is a \(\erre^m\)-valued Lagrange multiplier. The
  Fr\'echet derivative of \(\beta \mapsto F(\beta;\lambda)\) is given,
  in matrix notation, by
  \[
    D_\beta F(\beta;\lambda) = \beta - \alpha + L^\top \lambda.
  \]
  Setting \(D_\beta F(\beta;\lambda)\) equal to zero yields
  \(\beta(\lambda):= \alpha - L^\top \lambda\). Enforcing the
  constraint defining \(H\) implies
  \[
    v = L\beta = L\alpha - LL^\top \lambda,
  \]
  where \(LL^\top\) is invertible because \(L\) has full
  rank. Therefore, setting
  \[
    \lambda_\ast = \bigl( LL^\top \bigr)^{-1} (L\alpha - v)
  \]
  and
  \[
    \beta_\ast := \beta(\lambda_\ast)
    = \alpha - L^\top \bigl( LL^\top \bigr)^{-1} (L\alpha - v),
  \]
  it follows easily that \(\beta_\ast\) is the unique minimizer.
\end{proof}

For notational convenience, let \(h_k^{a,b}\) denote the function
\(h_k((\cdot-b)/a)\).  Let \(V_n\) denote the subspace of \(L^2\)
spanned by \(\bigl\{h_0^{a,b}, \ldots, h_n^{a,b}\bigr\}\), and
\[
  C_n := \Bigl\{
  \sum_{k=0}^n \alpha_k h_k^{a,b}: \,
  \alpha := (\alpha_0,\ldots,\alpha_n) \in \erre^{n+1},
  \, L\alpha = v \Bigr\}.
\]
It is clear that \(V_n\) is isomorphic to \(\erre^{n+1}\) and \(C_n\)
is a (closed) convex subset of \(V_n\).
We need to introduce some notation that is only needed because
\((h_k^{a,b})\) is an orthogonal basis of \(L^2\), but not
orthonormal. We shall set, for all \(k \in \bN\),
\[
  \check{\alpha}_k := \alpha_k {\norm{h_k^{a,b}}}_{L^2}, \qquad
  \hat{h}_k^{a,b} :=
  \frac{h_k^{a,b}}{{\norm{h_k^{a,b}}}_{L^2}},
\]
and similarly for other quantities. Moreover, the matrix
\(\hat{L} = (\hat{l}_{ij})\) is defined by
\[
\hat{l}_{ij} := \frac{l_{ij}}{{\norm{h_k^{a,b}}}_{L^2}}.
\]
Then \((\hat{\psi}_k^{a,b})\) is an orthonormal basis of \(L^2\), if
\(f = \sum_{k=0}^\infty \alpha_k h_k^{a,b}\) then
\(f = \sum_{k=0}^\infty \check{\alpha}_k \hat{h}_k^{a,b}\), and the
constraint \(L\alpha = v\) is equivalent to \(\hat{L}\check{\alpha} = v\).
\begin{prop}
  \label{prop:Hvinc}
  Let \(f \in L^2\), \((\alpha_k)\) the set of coefficients of its
  expansion with respect to the orthogonal basis of \(L^2\) formed by
  \((h_k^{a,b})\),
  \[
    \check{\beta} := \check{\alpha}
    - \hat{L}^\top \bigl( \hat{L}\hat{L}^\top \bigr)^{-1}
    (\hat{L}\check{\alpha} - v),
  \]
  and
  \[
    \beta_k := \frac{\check{\beta}_k}{{\norm{h_k^{a,b}}}_{L^2}}
    \qquad \forall k=1,\ldots,n.
  \]
  Then the projection of \(f \in L^2\) onto the convex set \(C_n\) is
  given by
  \[
    \widetilde{f}_n := \sum_{k=0}^n \beta_k h_k^{a,b}.
  \]
\end{prop}
\begin{proof}
  It is clear that \(\widetilde{f}_n \in C_n\), so it is enough to
  show that \(\norm{f - \widetilde{f}_n} \leq \norm{f - g}\) for every
  \(g \in C_n\). Suppose, by contradiction, that there exists
  \(\gamma \in \erre^{n+1}\) such that, setting
  \(g := \sum \gamma_k h_k^{a,b}\),
  \[
    \norm[\big]{f - g}_{L^2} < \norm[\big]{f - \widetilde{f}_n}_{L^2}.
  \]
  The inequality is equivalent to
  \begin{align*}
    &\norm[\bigg]{\sum_{k=0}^n (\check{\alpha}_k - \check{\gamma}_k)
      \hat{h}_k^{a,b}
      + \sum_{k=n+1}^\infty \check{\alpha}_k \hat{h}_k^{a,b}}^2_{L^2}\\
    &\hspace{3em}
  < \norm[\bigg]{\sum_{k=0}^n (\check{\alpha}_k - \check{\beta}_k)
      \hat{h}_k^{a,b}
      + \sum_{k=n+1}^\infty \check{\alpha}_k \hat{h}_k^{a,b}}^2_{L^2}.
  \end{align*}
  Therefore, by orthonormality,
  \[
    \norm[\big]{\check{\alpha} - \check{\gamma}}_{\erre^{n+1}}
    = \sum_{k=0}^n (\check{\alpha}_k - \check{\gamma}_k)^2
    < \sum_{k=0}^n (\check{\alpha}_k - \check{\beta}_k)^2
    = \norm[\big]{\check{\alpha} - \check{\beta}}_{\erre^{n+1}},
  \]
  which contradicts the optimality of \(\check{\beta}\), hence is
  impossible.
\end{proof}

It is now clear that Proposition \ref{prop:Hvinc} gives the solution
to the problem of finding the best approximation in \(L^2\) of \(f\)
among the linear combinations of Hermite functions with unit
integral. Explicit formulas for the coefficients of \(L\) (which
reduces to a vector in this case) can be found in \cite{cm:Herm}.  One
may also add as further constraint for the approximation
\(\widetilde{f}_n\) to satisfy the property
\begin{equation}
  \label{eq:AMP}
\int_\erre e^{ax+b} \widetilde{f}_n(x)\,dx = 1,
\end{equation}
that can be seen as an approximate martingale condition. Also in this
case explicit expressions for the corresponding matrix \(L\) are
available in \cite{cm:Herm}.

Once the desired, possibly constrained, approximation of \(f\) has
been obtained, it is easy to compute the \(L^2\) error. One has
\[
  \norm[\Big]{f - \sum_{k=0}^n \beta_k h^{a,b}_k}_{L^2}^2
  = {\norm{f}}_{L^2}^2 + \sum_{k=0}^n \beta_k^2 {\norm{h^{a,b}_k}}_{L^2}^2
  - 2 \sum_{k=0}^n \beta_k \ip{f}{h_k^{a,b}},
\]
where \({\norm{h^{a,b}_k}}_{L^2}^2 = a {\norm{h_k}}_{L^2}^2\) and
\(\ip{f}{h_k^{a,b}} = a {\norm{h_k}}_{L^2}^2 \alpha_k\), hence
\[
  \norm[\Big]{f - \sum_{k=0}^n \beta_k h^{a,b}_k}_{L^2}^2
  = {\norm{f}}_{L^2}^2 + a \Bigl( \sum_{k=0}^n \beta_k^2 {\norm{h_k}}_{L^2}^2
  - 2 \sum_{k=0}^n \alpha_k \beta_k {\norm{h_k}}_{L^2}^2 \Bigr).
\]
If \(\beta_k = \alpha_k\) for all \(k\), i.e. in the unconstrained
case, the squared \(L^2\) error simplifies to
\[
  {\norm{f}}_{L^2}^2 - a \sum_{k=0}^n \alpha_k^2 {\norm{h_k}}_{L^2}^2.
\]
Finally, recall that \({\norm{h_k}}_{L^2}^2 = \sqrt{\pi}\,k!\).

\begin{rmk}
  The approximation \(f_n\) is specified by \(n+3\) parameters, namely
  by the \(n+1\) coefficients \((\alpha_k)\) and the parameters
  \(a,b\). Constrained approximations of the type discussed above
  reduce the dimensionality by the number of independent constraints,
  hence the so-called approximate martingale approximation is
  specified by \(n+3-2=n+1\) parameters, as the constraints of having
  unit integral and integrating \(x \mapsto e^{ax+b}\) to one are
  independent (cf.~\cite{cm:Herm}).
\end{rmk}

\subsection{Variance-gamma densities}
\label{ssec:VGd}
Let us consider the Hermite approximation, in the sense of
\S\ref{ssec:Hlr}, of the density of a variance-gamma process \(X\) of
the type \(X_t = \theta\Gamma_t + \sigma W_{\Gamma_t} + \eta t\), as
defined in \S\ref{sec:VG}, so that, in particular, the process
\(\exp X\) is a \(\bQ\)-martingale. Let \(Y\) be the process defined
by \(Y_t := \theta\Gamma_t + \sigma W_{\Gamma_t}\) and, for a fixed
\(t \in \erre_+^\times\), denote the density of \(Y_t\) by \(f\). Then
the density of \(X_t\) is
\[
  f(x - \eta t) = \int_0^\infty \frac{1}{\sigma\sqrt{s}}
  \phi\biggl( \frac{x - \eta t - \theta\sqrt{s}}{\sigma\sqrt{s}}
  \biggr) \,d\lambda_{ct,\alpha}(s).
\]
If \(ct > 1/4\), which implies that \(f\), hence also
\(f(\cdot - \eta t)\), belong to \(L^2\), then, setting
\[
  \alpha_k = \frac{1}{k! \sqrt{\pi}}
  \int_\erre f(ax + b - \eta t) h_k(x)\,dx,
\]
one has, as an identity in \(L^2\),
\[
  f(x - \eta t) = \sum_{k=0}^\infty \alpha_k h_k\biggl(
  \frac{x-b}{a} \biggr).
\]
Since the density \(f\) admits an explicit expression (in terms of
special functions), the numerical computation of \((\alpha_k)\) is
easy to implement. The computation of \((\alpha_k)\) can also be
written as an integral with respect to a gamma maesure: as a first
step, note that, by Fubini's theorem,
\begin{align*}
  \alpha_k
  &\eqsim_k \int_\erre f(ax + b - \eta t) h_k(x)\,dx\\
  &\eqsim \int_\erre \int_{\erre_+} \frac{1}{\sigma\sqrt{s}}
  \phi\biggl( \frac{x - \eta t - \theta\sqrt{s}}{\sigma\sqrt{s}}
    \biggr) \,d\lambda_{ct,\alpha}(s) H_k\biggl( \sqrt{2}
    \frac{x-b}{a} \biggr) \phi\biggl(\frac{x-b}{a}\biggr)\,dx\\
  &= \int_{\erre_+} \frac{1}{\sigma\sqrt{s}}
    \int_\erre H_k\biggl( \sqrt{2} \frac{x-b}{a} \biggr)
    \phi\biggl( \frac{x-b}{a} \biggr)
    \phi\biggl( \frac{x - \eta t - \theta\sqrt{s}}{\sigma\sqrt{s}} \biggr)
    \,dx \,d\lambda_{ct,\alpha}(s).
\end{align*}
To continue we need an analytic lemma, the proof of which is
elementary, hence omitted.
\begin{lemma}
  Let \(\mu_1,\mu_2 \in \erre\),
  \(\sigma_1,\sigma_2 \in \erre_+^\times\), and
  \begin{align*}
    \mu &:= \mu_1 \frac{\sigma_2^2}{\sigma_1^2+\sigma_2^2}
          + \mu_2 \frac{\sigma_1^2}{\sigma_1^2+\sigma_2^2},\\
    \sigma &:= \frac{\sigma_1 \sigma_2}{\sqrt{\sigma_1^2+\sigma_2^2}},\\
    c &:= \frac{1}{\sigma^2} \biggl( \frac{\mu_1^2}{\sigma_1^2}
        + \frac{\mu_2^2}{\sigma_2^2} - \mu^2 \biggr).
  \end{align*}
  Then
  \[
    \phi\biggl( \frac{x-\mu_1}{\sigma_1} \biggr) %
    \phi\biggl( \frac{x-\mu_2}{\sigma_2} \biggr)
    = \frac{1}{\sqrt{2\pi}} \, \phi\biggl( \frac{x-\mu}{\sigma} \biggr)
      \, \exp c.
  \]
\end{lemma}
Therefore there exist functions
\(c, m, \varsigma\colon \erre_+ \to \erre\), depending on the parameters
\(a,b,\theta,\sigma\), such that
\[
  \phi\biggl( \frac{x-b}{a} \biggr) \phi\biggl(
  \frac{x - \eta t - \theta\sqrt{s}}{\sigma\sqrt{s}} \biggr)
  = e^{c(s)} \phi\biggl( \frac{x-m(s)}{\varsigma(s)} \biggr),
\]
hence the inner integral in the above expression for \(\alpha_k\) can
be written as
\[
  e^c \int_\erre H_k\biggl( \frac{x-b}{a/\sqrt{2}} \biggr)
  \phi\biggl( \frac{x - m}{\varsigma} \biggr)\,dx
  = \varsigma e^c \int_\erre H_k\biggl(
  \frac{\varsigma x + m - b}{a/\sqrt{2}} \biggr) \phi(x)\,dx,
\]
where the integral on the right-hand side can be reduced to computing
integrals of the type
\[
\int_\erre x^h e^{-x^2/2}\,dx, \qquad h \in \bN,
\]
that admit explicit expressions in terms of the gamma function. We
have thus shown that the function
\[
F(s) := \int_\erre H_k\biggl( \sqrt{2} \frac{x-b}{a} \biggr)
    \phi\biggl( \frac{x-b}{a} \biggr)
    \phi\biggl( \frac{x - \eta t - \theta\sqrt{s}}{\sigma\sqrt{s}}
    \biggr)\,dx
\]
can be written in terms of explicit functions (among which we include
the gamma function), hence that \(\alpha_k\) is obtained as the
integral of \(F\) with respect to the gamma measure
\(\lambda_{ct,\alpha}\).

\subsection{Heston densities}
\label{ssec:Hesd}
Let \(S\) be the discounted price process in a Heston stochastic
volatility model, as introduced in \S\ref{sec:Heston}, with the
normalizing assumptions \(q=1\) and \(S_0=1\), and let a strictly
positive time \(t\) be fixed. Assuming that the coefficients satisfy
the Feller condition, \(S_t\) is strictly positive with probability
one and, by Corollary \ref{cor:HL2}, \(\log S_t\) admits a density
\(f \in L^2\). Even though an explicit expression for the density
\(f\) is not available, the coefficients \((\alpha_k)\) of the Hermite
approximation of \(f\) can be written as integrals involving the
characteristic function \(\psi_t\) of \(\log S_t\), for which explicit
forms are known, and the Hermite functions \(h_k\), thanks to general
properties of the Fourier transform.

If \(g \in L^1\), let us define its Fourier transform \(\widehat{g}\)
by
\[
  \widehat{g}(\xi) := \frac{1}{\sqrt{2\pi}} \int_\erre
  g(x)e^{-ix\xi}\,dx.
\]
The same notation will be used for the usual extension of the Fourier
transform to \(L^2\). Moreover, recall that the Fourier transform is a
unitary operator of \(L^2\), i.e., for any \(g,h \in L^2\),
\[
  \ip[\big]{g}{h} = \ip[\big]{\overline{\widehat{g}}}{\widehat{h}},
\]
where \(\overline{\,\cdot\,}\) stands for complex conjugation.
\begin{prop}
  Let \(a \in \erre_+^\times\), \(b \in \erre\),
  \(\psi_{a,b}\colon \erre \to \bC\) the function defined by
  \[
    \psi_{a,b}(\xi) := \frac{e^{-ib\xi/a}}{a} \psi_t(\xi/a),
  \]
  and, for each \(k \in \bN\),
  \[
    \alpha_k :=
    \begin{cases}
      \displaystyle
      \frac{(-1)^{k/2}}{\pi k! \sqrt{2}}
        \ip{\operatorname{Re} \psi_{a,b}}{h_k},
      &\text{if } k \in 2\bN,\\[10pt]
      \displaystyle    
      \frac{(-1)^{k/2+3/2}}{\pi k! \sqrt{2}}
        \ip{\operatorname{Im} \psi_{a,b}}{h_k},
      &\text{if } k \in 2\bN + 1,
    \end{cases}
  \]
  Then the following identity holds in \(L^2\):
  \[
    f = \sum_{k=0}^\infty \alpha_k h_k\biggl(\frac{\cdot - b}{a} \biggr).
  \]
\end{prop}
\begin{proof}
  It follows immediately from \(f \in L^2\) that \(f_{a,b} \in L^2\),
  hence, as \(h_k \in L^2\) as well,
  \[
    \ip[\big]{f_{a,b}}{h_k} = \ip[\big]{\overline{\widehat{f_{a,b}}}}%
    {\widehat{h_k}},
  \]
  where, by well-known properties of the Fourier transform,  
  \[
    \widehat{f_{ab}}(\xi) = \frac{e^{ib\xi/a}}{a} \widehat{f}(\xi/a).
  \]
  Moreover, \href{https://en.wikipedia.org/wiki/Hermite_polynomials#Hermite_functions_as_eigenfunctions_of_the_Fourier_transform}{Hermite
    functions are eigenfunctions of the Fourier transform}, with
  \[
    \widehat{h_k}(\xi) = (-i)^k h_k(\xi) \qquad \forall k \in \bN.
  \]
  Since \(f\) is real, hence so is \(f_{a,b}\), one has
  \[
    \overline{\widehat{f_{a,b}}(\xi)} =\widehat{f_{a,b}}(-\xi)
  \]
  and, by definition of characteristic function (suppressing the index
  \(t\) for simplicity),
  \[
    \widehat{f}(\xi) = \frac{1}{\sqrt{2\pi}} \psi(-\xi),
  \]
  hence
  \[
    \overline{\widehat{f_{a,b}}(\xi)} = \frac{1}{a}
    \overline{e^{ib\xi/a} \widehat{f}(\xi/a)} = \frac{e^{-ib\xi/a}}{a}
    \overline{\widehat{f}(\xi/a)} = \frac{e^{-ib\xi/a}}{a\sqrt{2\pi}}
    \psi(\xi/a).
  \]
  Recalling the definition of \(\psi_{a,b}\), the above yields, after
  some algebraic manipulations,
  \[
    \ip{f_{a,b}}{h_k} = \frac{(-i)^k}{\sqrt{2\pi}} \ip{\psi_{a,b}}{h_k}
    =
    \begin{cases}
      \displaystyle
      \frac{(-1)^{k/2}}{\sqrt{2\pi}} \ip{\operatorname{Re} \psi_{a,b}}{h_k},
      &\text{if } k \in 2\bN,\\[10pt]
      \displaystyle    
      \frac{(-1)^{k+(k+1)/2}}{\sqrt{2\pi}} \ip{\operatorname{Im} \psi_{a,b}}{h_k},
      &\text{if } k \in 2\bN + 1,
    \end{cases}
  \]
  from which the claim follows immediately, recalling that \(\alpha_k
  = \ip{f_{a,b}}{h_k} / (k! \sqrt{\pi})\).
\end{proof}



\section{Empirical results}
\label{sec:res}
Approximations of variance-gamma and Heston densities are investigated
empirically in this section. As a first step we look at approximations
by Hermite methods of variance-gamma and Heston densities, in the
sense discussed in \S\ref{sec:Had}, in two test cases. We shall see
that, as far as the variance-gamma density is concerned, that is
determined by three parameters, estimates by Hermite approximations
with the same number of parameters seem rather poor. On the other
hand, Hermite approximations depending on five parameters of a Heston
density, that depends on the same number of parameters, appear to be
quite precise in terms of their \(L^2\) distance from the target,
although their pricing accuracy is not consistently satisfying.
Moreover, we conduct an extensive empirical analysis on a sample of
put option prices on the S\&P500 index for the whole year
2012.\footnote{Only options with low trading volumes or that violated
  price monotonicity with respect to strike prices were discarded.
  After these adjustments, the dataset included \(43\,469\)
  contracts. The most active day was December 21 (the 242nd day), with
  269 put prices across 14 expiration dates.}  Namely, we
compare the empirical performance of variance-gamma and Heston option
price estimators (in the sense of out-of-sample estimates) against
several classes of Hermite estimators depending on three and five free
parameters, respectively.

\subsection{Hermite approximation of variance-gamma densities}
\label{ssec:VGtoH}
\subsubsection{A synthetic example}
We first consider, as an example, a variance-gamma process \(Y\) with
parameters
\[
  \theta = 0.1, \quad \sigma = 0.3, \quad \alpha = 2,
\]
and we approximate the density \(f\) of its drifted version \(X\),
defined by \(X_t = Y_t+\eta t\), evaluated at \(t=1\), where the
notation of \S\ref{sec:VG} is used throughout. The random variable
\(X_t = X_1\) can be interpreted as \(\log S_t\), where \(S\) is a
martingale discounted price process of an asset, assuming no dividends
for simplicity. The mean and standard deviation of the density \(f\)
are \(-0.0505\) and \(0.308\), respectively, while its \(L^2\) norm is
\(1.01\).

In the notation and terminology of \S\ref{ssec:Hlr}, we are going to
consider four types of Hermite estimators, all depending on the same
number of free parameters as the variance-gamma distribution,
i.e. three. The simplest one assumes the parameters \(a\) and \(b\) to
satisfy \(b=-a^2/2\), with \(b\) equal to the mean of \(X_t\), and to
have two (unconstrained) Hermite coefficients \(\alpha_0, \alpha_1\).
As already mentioned in \S\ref{ssec:Hlr}, the resulting approximation
of \(f\), denoted by \(f_{1,p}\), can be interpreted as a perturbation
of the Black-Scholes model (hence the subscript \(p\)).  The
refinement of this approximation obtained by optimizing with respect
to the constant \(a\) is denoted by \(f^\ast_{1,p}\).
Just to fix notation, we set, for every \(n \in \bN\),
\[
  f_{n,p}(x) = \sum_{k=0}^n \alpha_k h_k\bigl( (x + a^2/2)/a \bigr),
\]
and \(f^\ast_{n,p}\) will stand for its optimized version (with
respect to \(a\)).
If the parameters \(a,b\) are chosen in terms of the density and
standard deviation of \(f\) as above, the Hermite approximation of
order three with the coefficients \((\alpha_k)\) satisfying the
so-called approximate martingale property has three free parameters,
and is thus another natural candidate to approximate the
variance-gamma density. Denoting the generic Hermite approximation of
order \(n \in \bN\) satisfying the approximate martingale property by
\(f_{n,m}\) and its optimized version (with respect to the parameter
\(a\)) by \(f^\ast_{n,m}\), we are thus going to consider \(f_{3,m}\)
and \(f^\ast_{3,m}\).
In the next subsection on the Hermite approximation of Heston
densities we shall also discuss approximations \(f_n\) with
independent parameters \(a,b\), but we do not consider them here. The
reason is that these approximations are characterized by \(n+3\)
parameters, hence we would have to take \(n=0\) to have three
parameters, but the \(f_0\) approximation is degenerate and is just
proportional to a Gaussian density.

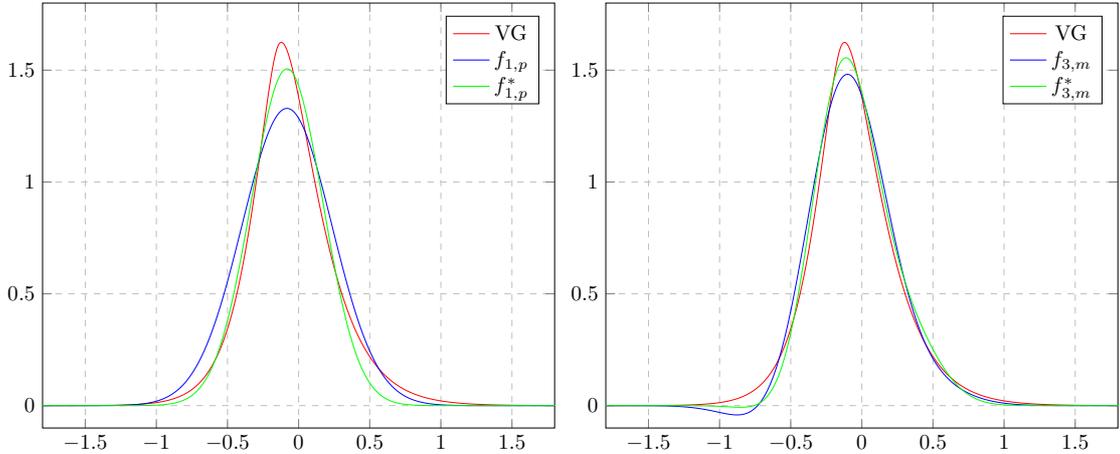
\begin{figure}[tbp]
  \centering
  \caption{Hermite approximations of a test variance-gamma (VG) density}

  \bigskip
  
  \begin{tikzpicture}[scale=0.8]
  \begin{axis}[
    xmin=-1.8, xmax=1.8,
    ymin=-0.1, ymax=1.8,
    legend pos=north east,
    xmajorgrids=true,
    ymajorgrids=true,
    grid style=dashed, ]

    \addplot[red] table [x=x,y=f] {figVG1.dat};
    \addlegendentry{VG}

    \addplot[blue] table [x=x,y=fHBS] {figVG1.dat};
    \addlegendentry{\(f_{1,p}\)}
    
    \addplot[green] table [x=x,y=fHBSopt] {figVG1.dat};
    \addlegendentry{\(f^\ast_{1,p}\)}
  \end{axis}
  \end{tikzpicture}
  \begin{tikzpicture}[scale=0.8]
  \begin{axis}[
    xmin=-1.8, xmax=1.8,
    ymin=-0.1, ymax=1.8,
    legend pos=north east,
    xmajorgrids=true,
    ymajorgrids=true,
    grid style=dashed, ]

    \addplot[red] table [x=x,y=f] {figVG2.dat};
    \addlegendentry{VG}

    \addplot[blue] table [x=x,y=fH_AMP] {figVG2.dat};
    \addlegendentry{\(f_{3,m}\)}
    
    \addplot[green] table [x=x,y=fH_AMPopt] {figVG2.dat};
    \addlegendentry{\(f^\ast_{3,m}\)}
  \end{axis}
  \end{tikzpicture}
  \label{fig:VG1}
\end{figure}

The distance from \(f\) of \(f_{1,p}\), \(f_{3,m}\), and of their starred
versions in the sense of the \(L^2\), \(L^1\), and \(L^\infty\) norms,
expressed in percentage of the corresponding norm of \(f\), are
collected in Table \ref{tab:VG1}.
\begin{table}[h]
  \renewcommand{\arraystretch}{1.3} 
  \setlength{\tabcolsep}{8pt} 
  \centering
  \caption{Hermite Approximation Error of a Test Variance-Gamma Density}    

  \medskip

  \parbox{0.9\textwidth}{\small For all
    \(g \in \{f_{1,p}, f^\ast_{1,p}, f_{3,m}, f^\ast_{3,m}\}\) and
    \(q \in \{1, 2, \infty\}\), the relative error
    \({\norm{f-g}}_{L^q} / {\norm{f}}_{L^q}\) in percentage points is
    reported, where \(f\) is the test variance-gamma density.}
  \label{tab:VG1}

  \bigskip

  \begin{tabular}{c|cccc}
    \hline
    Norm & \(f_{1,p}\) & \(f^\ast_{1,p}\) & \(f_{3,m}\) & \(f^\ast_{3,m}\) \\
    \hline
    \(L^2\)     & 17.6 & 9.87 & 10.4 & 7.39\\
    \(L^1\)     & 19.7 & 12.4 & 12.6 & 9.64\\
    \(L^\infty\) & 19.0 & 9.25 & 10.1 & 5.14\\
    \hline
  \end{tabular}
\end{table}
As can easily be inferred from the reported values (cf. also Figure
\ref{fig:VG1}), approximation errors appear too large to be useful for
the purpose of approximating integrals against \(f\) by integrals
against any one of its approximations, as would be the case in option
pricing. However, it is at the same time remarkable that a function
defined in terms of the standard deviaton of \(f\) and two Hermite
polynomials, hence a largely uninformed device, can ``explain'' 90\%
of the \(L^2\) norm of \(f\). The relative \(L^\infty\) error achieved
by the \(f^\ast_{3,m}\) is also remarkable.

\medskip

To investigate how the various approximation schemes perform as
estimators of option prices, we proceed as follows: let
\(K_1,\ldots,K_N\) be strike prices randomly sampled from a uniform
distribution on \([1/2,5/4]\) with \(N=20\). Variance-gamma prices of
put options with this set of strike prices can then be computed. That
is, assuming that the underlying is an asset with discounted price
process modeled by \(S = \exp X\), i.e. by a variance-gamma dynamics
with the parameters introduced at the beginning of the subsection,
corresponding put option prices are obtained. Each approximation of
\(f\) produces an estimate of the variance-gamma prices of the put
options, e.g. \(f_{1,p}\) determines the approximation
\[
\int_\erre (K_i - e^x)^+ f_{1,p}(x)\,dx, \qquad i=1,\ldots,N.
\]
The absolute relative pricing error (an element of \([0,1]^N\)) for
each of the proposed approximations is then determined, and
descriptive statistics are collected in Table \ref{tab:VG2}.
\begin{table}[t]
  \centering
  \renewcommand{\arraystretch}{1.3} 
  \setlength{\tabcolsep}{10pt}
  \caption{Absolute Relative Pricing Errors by Hermite Approximations
    of a Test Variance-Gamma Density}

  \medskip

  \parbox{0.9\textwidth}{\small For a set of twenty randomly generated
    strike prices in the interval \([0.5,1.25]\), descriptive
    statistics on the absolute relative error of put option prices
    generated by the approximations
    \(\{f_{1,p}, f^\ast_{1,p}, f_{3,m}, f^\ast_{3,m}\}\) with respect
    to the put option prices generated by the test variance-gamma
    density \(f\) are reported (all data in percentage points).}
  \label{tab:VG2}

  \bigskip

  \begin{tabular}{ccccc}
    \multicolumn{5}{c}{\textsc{Mean}} \\
    \hline
    & \(f_{1,p}\) & \(f^\ast_{1,p}\) & \(f_{3,m}\) & \(f^\ast_{3,m}\) \\  
    \hline
    &33.0 & 8.17 & 100 & 20.8 \\
    \hline
  \end{tabular}

  \bigskip 

  \begin{tabular}{c|cccc}
    \multicolumn{5}{c}{\textsc{Distribution}} \\
    \hline
    Quantile & \(f_{1,p}\) & \(f^\ast_{1,p}\) & \(f_{3,m}\) & \(f^\ast_{3,m}\) \\
    \hline
    25\% & 14.0  &  2.17 & 100 &  5.88 \\
    50\% & 31.2  &  2.84 & 100 & 12.1  \\
    75\% & 43.3  &  3.54 & 100 & 18.0  \\
    95\% & 74.1  & 45.9  & 100 & 82.1  \\
    99\% & 74.7  & 67.1  & 100 & 100   \\
    \hline
  \end{tabular}
\end{table}
It turns out that, in spite of the relatively poor approximations of
\(f\) achieved by the Hermite approximations, the option price
estimator associated to \(f^\ast_{1,p}\) obtains quite good results,
with 75\% of the price estimates within 3.54\% of the target
variance-gamma price.

We also consider ``corrected'' option prices, in the sense of
\eqref{eq:pin}, where we assume that the variance-gamma price of the
put option with strike price \(k_0\) is known. Of course this reduces
the size of the sample from \(N\) to \(N-1\). For simplicity, we take
\(k_0\) equal to the median of the \(N\) strike prices
\(K_1,\ldots,K_N\). Descriprive statistics of the absolute relative
error in percentage points are collected in Table \ref{tab:VG3}.
\begin{table}[t]
  \renewcommand{\arraystretch}{1.3} 
  \setlength{\tabcolsep}{10pt}
  \centering
  \caption{Absolute Relative Pricing Errors by Hermite Approximations
    of a Test Variance-Gamma Density (``Corrected'' Prices)}

  \medskip

  \parbox{0.9\textwidth}{\small For the same set of twenty randomly
    generated strike prices in the interval \([0.5,1.25]\) as in Table
    \ref{tab:VG2}, descriptive statistics on the absolute relative
    error of put option prices generated by the approximations
    \(f_{1,p}, f^\ast_{1,p}, f_{3,m}, f^\ast_{3,m}\) with respect
    to the ``corrected'' put option prices, in the sense of
    \eqref{eq:pin}, generated by the test variance-gamma density \(f\)
    are reported (all data in percentage points).}
  \label{tab:VG3}

  \bigskip

  \begin{tabular}{ccccc}
    \multicolumn{5}{c}{\textsc{Mean}} \\
    \hline
    & \(f_{1,p}\) & \(f^\ast_{1,p}\) & \(f_{3,m}\) & \(f^\ast_{3,m}\) \\  
    \hline
    & 18.0 & 2.57 & 91.3 & 17.1 \\
    \hline
  \end{tabular}

  \bigskip 

  \begin{tabular}{c|cccc}
    \multicolumn{5}{c}{\textsc{Distribution}} \\
    \hline
    Quantile & \(f_{1,p}\) & \(f^\ast_{1,p}\) & \(f_{3,m}\) & \(f^\ast_{3,m}\) \\
    \hline
    25\% & 0.124 & 0.586 & 100 & 0.276 \\
    50\% & 0.343 & 0.837 & 100 & 0.462 \\
    75\% & 4.17  & 1.00 & 100 & 0.533 \\
    95\% & 100 & 16.7  & 100 & 100   \\
    99\% & 100 & 24.8  & 100 & 100   \\
    \hline
  \end{tabular}
\end{table}
In this case, apart from the \(f_{3,m}\) estimator, that is
consistently off target, all estimators perform reasonably well, with
occasional very large errors compromising their average error in some
cases. In particular, the \(f^\ast_{1,p}\) and \(f^\ast_{3,m}\)
estimators stand out, with 75\% of their price estimates within
4.17\% and 0.533\% of the target price, respectively.

\medskip

Lastly, we address a different but related question: given the above
set of option prices generated by a variance-gamma model, does a
Hermite option price estimator calibrated to these prices display
acceptable accuracy?
Numerical results suggest that calibrating to option prices an
approximation to \(f\) of the type \(f_{1,p}\) (this calibration was
called \(\mathrm{H}_\sigma\) in \cite{cm:Herm}) produces quite
satisfying results, with mean relative pricing error equal to 4.67\%,
and quantiles at 75\% and 95\% level equal to 4.32\% and 14.7\%,
respectively.

\subsubsection{One day of real data}
As a next step, we consider one day of real data (day 242 of our
dataset), made up by several blocks of options having the same time to
maturity. The parameters of a variance-gamma density are then
calibrated to each block, minimizing the mean absolute relative
pricing error.
Errors and values of the calibrated parameters for each block are
displayed in Table \ref{tab:VG4}.
\begin{table}[t]
  \renewcommand{\arraystretch}{1.3} 
  \setlength{\tabcolsep}{8pt} 
  \centering
  \caption{Variance-Gamma Calibration on One Day of Data}

  \medskip
  
  \parbox{0.9\textwidth}{\small Mean absolute relative pricing error
    for each block of day 242 of the dataset. Each row corresponds to
    a block and contains the time to maturity in days, the error in
    percentage points, and the values of the calibrated parameters.}
  \label{tab:VG4}

  \bigskip

  \begin{tabular}{c|c|r|r|r}
    \hline
    Maturity & Error & \(\theta\) & \(\sigma\) & \(\alpha\) \\
    \hline
    4   & 21.4  & -0.0743  & 0.147  & 0.743 \\
    10  & 15.2  & -0.0910  & 0.175  & 2.800 \\
    13  & 15.5  & -0.0667  & 0.180  & 2.980 \\
    17  & 3.61  & -0.0209  & 0.177  & 8.590 \\
    24  & 2.37  & -0.0046  & 0.175  & 6.610 \\
    32  & 6.55  & -0.0405  & 0.161  & 3.760 \\
    60  & 3.65  &  0.0292  & 0.193  & 2.330 \\
    88  & 3.61  & -0.0562  & 0.176  & 2.510 \\
    186 & 11.0  & -0.1670  & 0.134  & 1.660 \\
    277 & 0.332 &  0.0667  & 0.234  & 0.788 \\
    368 & 0.888 & -0.2020  & 0.078  & 1.040 \\
    732 & 15.4  & -0.1400  & 0.090  & 0.272 \\
    \hline
  \end{tabular}
\end{table}
For each block, we approximate the calibrated variance-gamma density
using the Hermite schemes introduced above, and compute their
estimation errors in terms of the norms \(L^2\), \(L^1\), and
\(L^\infty\) (except for the first three blocks, for which the
condition for the density to be in \(L^2\) is not satisfied). The
\(L^2\) estimation errors in the remaining nine blocks range between
28\% and 62\%, thus making the corresponding approximations of very
limited use. The calibration of Hermite estimators to the option
prices corresponding to the calibrated variance-gamma models
unfortunately produces comparable results, leading to the conclusion
that Hermite approximation of variance-gamma densities and of option
prices generated by such densities may not be satisfactory, at least
not when the Hermite approximation schemes depend only on three free
parameters.

\subsection{Hermite approximation of Heston densities}
\label{ssec:HestoH}
\subsubsection{A synthetic example}
As an example we consider, in the notation of \S\ref{sec:Heston}, the
density \(f\) at time \(t=1\) of \(\log S_t\), assuming \(S_0=1\),
\(q=0\), and the parameters
\[
v_0 = 0.05, \quad \kappa = 1, \quad \theta = 0.1, \quad
\eta = 0.25, \quad \rho = -0.75,
\]
that satisfy the Feller condition. The density \(f\) has mean and
standard deviation equal to \(-0.0342\) and \(0.271\), respectively,
and its \(L^2\) norm is \(1.11\).

In analogy to the discussion in \S\ref{ssec:VGtoH}, we consider Hermite
approximations of \(f\) that depends on the same number of parameters
specifying the Heston density, that is five. The simplest Hermite
approximation is thus \(f_{3,p}\), defined by
\[
  f_{3,p}(x) = \sum_{k=0}^3 \alpha_k h_k\bigl( (x + a^2/2)/a \bigr),
\]
where the parameter \(b\) is equal to the mean of \(\log S_t\) and
\(a\) is such that \(b=-a^2/2\).  The refined approximation
\(f^\ast_{3,p}\) obtained by optimizing with respect to the parameter
\(a\) is also employed. The relative \(L^2\), \(L^1\), and
\(L^\infty\) errors in the approximation of \(f\) are reported in the
first two columns of Table \ref{tab:Hes1}.
\begin{table}[t]
  \renewcommand{\arraystretch}{1.3} 
  \setlength{\tabcolsep}{8pt}
  \centering  
  \caption{Hermite Approximation Errors of a Test Heston Density}

  \medskip

  \parbox{0.9\textwidth}{\small For all
    \(g \in \{f_{3,p}, f^\ast_{3,p}, f_2, f^\ast_2, f_{5,m},
    f^\ast_{5,m}\}\) and \(q \in \{1, 2, \infty\}\), the relative
    error \({\norm{f-g}}_{L^q} / {\norm{f}}_{L^q}\) in percentage points
    is reported, where \(f\) is the test Heston density.}
  \label{tab:Hes1}

  \bigskip

  \begin{tabular}{c|cccccc}
    \hline
    Norm & \(f_{3,p}\) & \(f^\ast_{3,p}\) & \(f_2\) & \(f_2^\ast\) & \(f_{5,m}\) & \(f^\ast_{5,m}\) \\
    \hline
    \(L^2\)       & 3.53 & 3.08 & 12.2 & 7.17 & 3.07 & 2.60 \\
    \(L^1\)       & 4.68 & 4.10 & 15.6 & 10.5 & 6.04 & 3.55 \\
    \(L^\infty\)  & 3.35 & 2.83 & 11.2 & 5.18 & 6.63 & 1.62 \\
    \hline
  \end{tabular}
\end{table}
Both approximations appear to be satisfactory: the \(L^2\) and
\(L^\infty\) do not exceed approximately \(3.5\%\), while the \(L^1\)
error of \(f_{3,p}\) is approximately \(4.7\%\), improving to
\(4.1\%\) for \(f_{3,p}^\ast\). Moreover, note that the improvement in
the approximation caused by the optimization with respect to the
parameter \(a\) is significant but not dramatic. This observation
confirms that interpreting Hermite approximations as perturbations of
the Black-Scholes model is at least meaningful.
As a further Hermite approximation of \(f\) depending on five
parameters, we consider the approximation of order two with
uncontrastrained constants \(a\) and \(b\), that is, in the notation
\S\ref{ssec:Hlr}, \(f_2\), as well as its refinement \(f_2^\ast\)
obtained by optimizing with respect to \(a\) and \(b\). The
approximation errors of these two schemes (see the third and fourth
columns of Table \ref{tab:Hes1}) are much higher than those attained
by \(f_{3,p}\) and \(f^\ast_{3,p}\), suggesting that it is not
efficient to swap a location-scale parameter with an order of Hermite
expansion. Moreover, even though the optimized approximation
\(f_2^\ast\) is much better than the approximation \(f_2\), the former
assumes negative values on a non-negligible portion of the right tail,
which is certainly an undesirable feature.
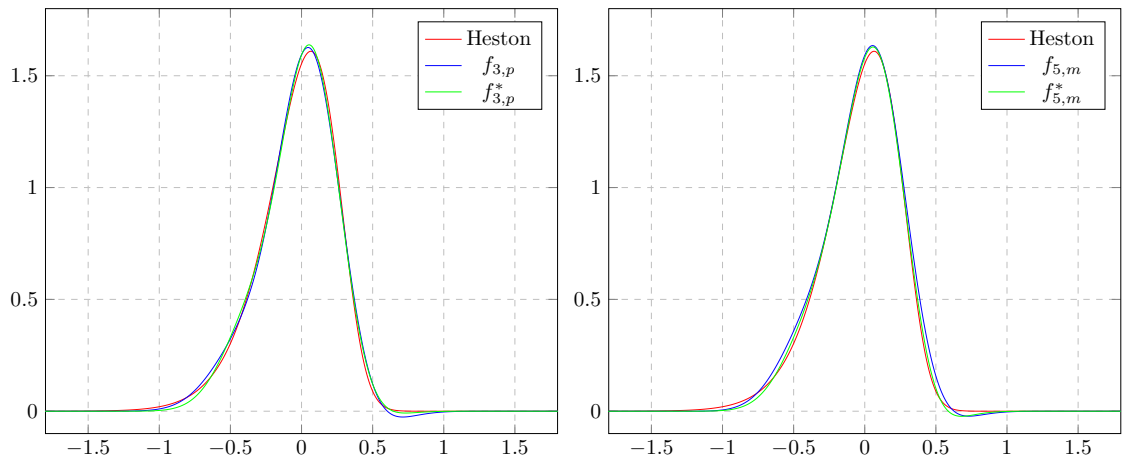
\begin{figure}[tbp]
  \centering
  \caption{Hermite approximations of a test Heston density}

  \bigskip
  
  \begin{tikzpicture}[scale=0.8]
  \begin{axis}[
    xmin=-1.8, xmax=1.8,
    ymin=-0.1, ymax=1.8,
    legend pos=north east,
    xmajorgrids=true,
    ymajorgrids=true,
    grid style=dashed, ]

    \addplot[red] table [x=x,y=f] {figHeston1.dat};
    \addlegendentry{Heston}

    \addplot[blue] table [x=x,y=fHBS] {figHeston1.dat};
    \addlegendentry{\(f_{3,p}\)}
    
    \addplot[green] table [x=x,y=fHBSopt] {figHeston1.dat};
    \addlegendentry{\(f^\ast_{3,p}\)}
  \end{axis}
  \end{tikzpicture}
  \begin{tikzpicture}[scale=0.8]
  \begin{axis}[
    xmin=-1.8, xmax=1.8,
    ymin=-0.1, ymax=1.8,
    legend pos=north east,
    xmajorgrids=true,
    ymajorgrids=true,
    grid style=dashed, ]

    \addplot[red] table [x=x,y=f] {figHeston2.dat};
    \addlegendentry{Heston}

    \addplot[blue] table [x=x,y=fH_AMP] {figHeston2.dat};
    \addlegendentry{\(f_{5,m}\)}
    
    \addplot[green] table [x=x,y=fH_AMPopt] {figHeston2.dat};
    \addlegendentry{\(f^\ast_{5,m}\)}
  \end{axis}
  \end{tikzpicture}
  \label{fig:Hes1}
\end{figure}
The last pair of Hermite approximations we consider is composed of
\(f_{5,m}\) and \(f^\ast_{5,m}\). Recall that the former is obtained
constraining the Hermite coefficients \((\alpha_k)\) of \(f_5\), with
\(b=-a^2/2\), to satisfy the so-called approximate martingale
property, and the latter is its optimized version (with respect to
\(a\)). The approximation errors are collected in the last two columns
of Table \ref{tab:Hes1}. While the approximation \(f_{5,m}\) has
essentially the same \(L^2\) error of \(f^\ast_{3,p}\), but
significantly higher \(L^1\) and \(L^\infty\) errors also than those
of the more elementary \(f_{3,p}\), the optimized approximation
\(f^\ast_{5,m}\) has significantly lower errors than all other
approximations, with an impressive relative \(L^\infty\) error of
\(1.62\%\) (cf. Figure \ref{fig:Hes1}).

\medskip

The good accuracy of several approximations of \(f\) introduced thus
far in terms of \(L^2\) (as well as \(L^1\) and \(L^\infty\)) norms
does not necessarily translate in good estimates of option
prices. Moreover, if \(g_1,g_2\) are approximations of \(f\) and the
distance of \(g_1\) from \(f\) is less than the distance of \(g_2\)
from \(f\) in any one of the above norms, it does not follow that
approximate option prices computed replacing \(f\) by \(g_1\) are
better than those computed replacing \(f\) by \(g_2\). Therefore, in
analogy to what we have done in \S\ref{ssec:VGtoH}, we consider the
same set \(K_1,\ldots,K_N\) of strike prices, for which prices of put
options are computed assuming that the underlying is an asset with
discounted asset price modeled by \(S\), i.e. by a Heston dynamics
with the parameters introduced at the beginning of the subsection.
These Heston prices are then compared to the price estimates produced
by each one of the Hermite approximations considered above. The mean
absolute relative pricing errors and quantiles of their empirical
distributions are reported in Table \ref{tab:Hes2}.
\begin{table}[tbp]
  \renewcommand{\arraystretch}{1.3} 
  \setlength{\tabcolsep}{10pt}
  \centering
  \caption{Errors of Hermite Estimates of Synthetic Heston Prices}
  \label{tab:Hes2}

  \medskip

  \parbox{0.9\textwidth}{\small For the same set of twenty randomly
    generated strike prices in the interval \([0.5,1.25]\) as in Table
    \ref{tab:VG2}, descriptive statistics on the absolute relative
    error of put option prices generated by the approximations
    \(\{f_{3,p}, f^\ast_{3,p}, f_2, f^\ast_2, f_{5,m},
    f^\ast_{5,m}\}\) with respect to the put option prices generated
    by the test Heston density \(f\) are reported (all data in
    percentage points).}
  
  \bigskip

  \begin{tabular}{cccccc}
    \multicolumn{6}{c}{\textsc{Mean}}\\
    \hline
    \(f_{3,p}\) & \(f^\ast_{3,p}\) & \(f_2\) & \(f_2^\ast\) & \(f_{5,m}\)
    & \(f^\ast_{5,m}\)\\
    \hline
    0.62 & 4.95 & 20.5 & 17.2 & 73.0 & 4.02 \\
    \hline
  \end{tabular}

  \bigskip 

  \begin{tabular}{c|cccccc}
    \multicolumn{7}{c}{\textsc{Distribution}}\\
    \hline
    Quantile & \(f_{3,p}\) & \(f^\ast_{3,p}\) & \(f_2\) & \(f_2^\ast\) & \(f_{5,m}\) & \(f^\ast_{5,m}\) \\
    \hline
    25\% & 0.0523 & 2.91 & 12.1 & 10.5 & 75.3 & 1.84 \\
    50\% & 0.382  & 3.58 & 14.7 & 12.7 & 84.2 & 2.77 \\
    75\% & 1.05   & 7.55 & 32.6 & 26.9 & 94.2 & 6.42 \\
    95\% & 2.06   & 10.9 & 45.7 & 37.1 & 95.8 & 10.3 \\
    99\% & 2.92   & 13.1 & 53.0 & 42.8 & 96.2 & 12.7 \\
    \hline
  \end{tabular}
\end{table}
The basic \(f_{3,p}\) approximation, that, we recall, can be
interpreted as a perturbation of the Black-Scholes model, clearly
outperforms all other approximations, and appears to be a good
estimator of Heston option prices, with all relative errors not
exceeding \(3\%\). It is interesting to observe that the optimized
approximation \(f^\ast_{3,p}\) performs much worse than its basic
version, and that the \(f_{5.m}\) approximation is completely off
target, while its optimized version performs reasonably well, but
still considerably worse than the more elementary \(f_{3,p}\)
approximation. Hence the added level of sophistication of
\(f_{5,m}^\ast\) may not be worth the effort, at least as far as the
problem of estimating the price of a limited set of put options is
concerned.  Moreover, the rule of thumb according to which swapping a
Hermite function for a degree of freedom in the choice of parameters
\(a,b\) is inefficient is confirmed by these data.

Considering the price of the put option with the median strike price
as observed, hence known, and applying \eqref{eq:pin}, we obtain
``corrected'' Hermite price estimates, for which we table mean
absolute relative error and quantiles of its empirical distribution in
Table \ref{tab:Hes3}.
\begin{table}[tbp]
  \renewcommand{\arraystretch}{1.3} 
  \setlength{\tabcolsep}{10pt}
  \centering
  \caption{Absolute Relative Pricing Errors by Hermite Approximations
    of a Test Heston Density (``Corrected'' Prices)}

  \medskip

  \parbox{0.9\textwidth}{\small For the same set of twenty randomly
    generated strike prices in the interval \([0.5,1.25]\) as in Table
    \ref{tab:VG2}, descriptive statistics on the absolute relative
    error of put option prices generated by the approximations
    \(f_{3,p}, f^\ast_{3,p}, f_2, f^\ast_2, f_{5,m}, f^\ast_{5,m}\)
    with respect to the ``corrected'' put option prices, in the sense
    of \eqref{eq:pin}, generated by the test Heston density \(f\) are
    reported (all data in percentage points).}
  \label{tab:Hes3}

  \bigskip

  \begin{tabular}{cccccc}
    \multicolumn{6}{c}{\textsc{Mean}} \\
    \hline
    \(f_{3,p}\) & \(f^\ast_{3,p}\) & \(f_2\) & \(f^\ast_2\) & \(f_{5,m}\) & \(f^\ast_{5,m}\) \\
    \hline
    0.270 & 0.729 & 7.09 & 6.41 & 81.3 & 0.187 \\
    \hline
  \end{tabular}

  \bigskip 

  \begin{tabular}{c|cccccc}
    \multicolumn{7}{c}{\textsc{Distribution}} \\
    \hline
    Quantile & \(f_{3,p}\) & \(f^\ast_{3,p}\) & \(f_2\) & \(f^\ast_2\) & \(f_{5,m}\) & \(f^\ast_{5,m}\) \\
    \hline
    25\% & 0.140 & 0.218 & 0.779 & 1.17 & 38.2  & 0.00215 \\
    50\% & 0.308 & 0.284 & 1.09  & 1.34 & 77.2  & 0.176   \\
    75\% & 0.337 & 0.658 & 1.98  & 2.16 & 100   & 0.371   \\
    95\% & 0.729 & 4.32  & 52.1  & 44.4 & 100   & 0.425   \\
    99\% & 1.04  & 5.78  & 53.7  & 46.4 & 100   & 0.435   \\
    \hline
  \end{tabular}
\end{table}
While the overall good accuracy of \(f_{3,p}\) is confirmed, the
picture regarding \(f^\ast_{5,m}\), with a relative pricing error not
exceeding approximately \(0.44\%\), improves drastically, suggesting
that, under the realistic assumption that a ``true'' option price is
available, the added level of sophistication of \(f_{5,m}^\ast\)
compared to the more elementary \(f_{3,p}\) may indeed pay off.

\medskip

To conclude the analysis, we investigate how Hermite price estimators
calibrated to the above set of Heston option prices perform. We shall
only consider a Hermite estimator of the type \(f_{3,p}\), about which
the numerical results suggest that it matches the Heston prices with a
relative pricing error that is essentially negligible: the
mean absolute relative error is \(0.0538\%\), the median error is
\(0.0397\%\), and the largest error is \(0.179\%\).
While it is clear that calibrating a Heston estimator to observed
prices must produce better estimates than going through a direct
approximation of the density \(f\) (simply by looking at the objective
functions of the two procedures), the achieved accuracy is nonetheless
remarkable. One should also mention that the calibration of a Hermite
estimator of put option prices does not ``care'' about the density of
the underlying on the set
\([\log K_{\mathrm{max}},+\infty\mathclose[\), where
\(K_{\mathrm{max}}\) is the largest strike price in the set of
observed options. In this sense, the task of calibrating a Hermite
price estimator to a set of put options is less demanding than
approximating a density. Of course the same remark would apply,
\emph{mutatis mutandis}, to a set of call options.

\subsubsection{One day of real data}
We are now going to consider issues analogous to those treated in the
previous paragraph when applied to the data pertaining to day 242 of
our dataset.
A Heston density is calibrated to each block, minimizing the
\(\ell^1\) norm of relative pricing errors. The estimated Heston
parameters for each block, see Table \ref{tab:Hes4}, display some
patterns: the initial value of the squared volatility \(v_0\) is close
to zero; the long-term equilibrium level \(\theta\) of the squared
volatility exhibit some variation, ranging from approximately \(0.08\)
to \(3.94\), but taking most often values between \(0.1\) and \(0.2\);
the correlation \(\rho\) between the driving Wiener processes is
negative and very large in absolute value (except for one case),
ranging from \(-0.85\) to \(-0.98\); the speed of mean reversion
\(\kappa\) and the volality \(\eta\) of the squared volatility, that
also vary in a wide range, exhibit some degree of correlation between
each other: small values of \(\kappa\) tend to be associated with
small values of \(\eta\).
\begin{table}[tbp]
  \renewcommand{\arraystretch}{1.3} 
  \setlength{\tabcolsep}{10pt}
  \centering  
  \caption{Heston Calibration on One Day of Data}

  \medskip
  
  \parbox{0.9\textwidth}{\small Mean absolute relative pricing error
    for each block of day 242 of the dataset. Each row corresponds to
    a block and contains the time to maturity in days, the error in
    percentage points, and the values of the calibrated parameters.}
  \label{tab:Hes4}

  \bigskip
  \begin{tabular}{c|c|r|r|r|r|r}
    \hline
    Maturity & Error & \(v_0\) & \(\theta\) & \(\rho\) & \(\kappa\)
    & \(\eta\) \\
    \hline
    4 & 51.7 & 0.000435 & 3.94 & -0.984 & 0.71 & 2.37\\
    10 & 31.6 & 0.000542 & 0.125 & -0.979 & 14.2 & 1.88\\
    13 & 26.5 & 1.55E-09 & 0.518 & -0.982 & 2.24 & 1.52\\
    17 & 5.29 & 0.0024 & 0.0822 & -0.966 & 11.7 & 1.39\\
    24 & 4.78 & 0.00266 & 0.192 & -0.925 & 3.17 & 1.1\\
    32 & 14.7 & 1.79E-07 & 0.175 & -0.971 & 2.94 & 1.01\\
    60 & 10.3 & 0.00211 & 0.162 & -0.941 & 1.83 & 0.769\\
    88 & 7.02 & 0.00693 & 2.05 & -0.854 & 0.0813 & 0.577\\
    186 & 11.6 & 0.00144 & 0.14 & -0.918 & 1.02 & 0.533\\
    277 & 20.1 & 0.0181 & 1.13 & 0.929 & 0.0723 & 2.78E-05\\
    368 & 0.773 & 0.00205 & 0.141 & -0.935 & 0.681 & 0.438\\
    732 & 19.7 & 0.00206 & 0.356 & -0.952 & 0.163 & 0.341\\
    \hline
  \end{tabular}
\end{table}
On each block we approximate the calibrated Heston density \(f\) by
finite linear combinations of scaled and shifted Hermite functions,
and compute their approximation errors in the norms \(L^2\), \(L^1\),
and \(L^\infty\), that are reported (in percentage terms of the
corresponding norm of \(f\)) in Table \ref{tab:Hes5}.
\begin{table}[tbp]
  \renewcommand{\arraystretch}{1.3} 
  \setlength{\tabcolsep}{12pt}
  \centering
  \caption{Hermite Approximation Errors of Calibrated Heston Densities
    in One Day of Real Data}

  \medskip

  \parbox{0.9\textwidth}{\small For all
    \(g \in \{f_{3,p}, f^\ast_{3,p}, f_{5,m}, f^\ast_{5,m}\}\) and
    \(q \in \{1, 2, \infty\}\), the average relative error
    \({\norm{f-g}}_{L^q} / {\norm{f}}_{L^q}\) in percentage points is
    reported, where \(f\) is the calibrated Heston density for each
    block of day 242. The average is taken over all blocks.}
  \label{tab:Hes5}

  \bigskip
  
  \begin{tabular}{c|cccc}
    \hline
    Norm & \(f_{3,p}\) & \(f^{\ast}_{3,p}\) & \(f_{5,m}\) & \(f^\ast_{5,m}\) \\
    \hline
    \(L^{2}\)    & 26.4 & 15.6 & 23.5 & 15.1 \\
    \(L^{1}\)    & 35.1 & 23.2 & 32.2 & 22.3 \\
    \(L^\infty\) & 27.3 & 14.3 & 23.8 & 13.0 \\
    \hline
  \end{tabular}
\end{table}
Errors are much higher than those obtained by the same methods for the
test Heston density considered above. A tentative explanation for this
phenomenon, guided by inspecting closely a few cases, is that
calibrated Heston densities tend to be peaked around zero, and for
such functions Hermite approximations converge slowly in \(L^2\).
Moreover, the approximations can display oscillations.
This gloomy picture is completely changed if one calibrates Hermite
estimators to the option prices implied by the calibrated Heston
densities. Mean absolute errors of Hermite option prices estimates
with respect to the implied Heston prices for each block are reported
in Table \ref{tab:Hes6}.
\begin{table}[tbp]
  \renewcommand{\arraystretch}{1.3} 
  \setlength{\tabcolsep}{10pt}
  \centering
  \caption{Mean Absolute Relative Errors of Hermite Prices Calibrated
    on Heston Prices}
  \medskip

  \parbox{0.9\textwidth}{\small For each block of put options of day
    242 with the same time to maturity, the Hermite approximation
    \(f^\ast_{3,p}\) is calibrated to the option prices implied by the
    Heston calibration, and the mean absolute relative error is
    reported. Each row refers to a block, where the first column
    contains the time to maturity in days, the second column the
    number of different strikes, and the third column the error in
    percentage points.}
  \label{tab:Hes6}

  \bigskip

  \begin{tabular}{ccc}
    \hline
    Maturity & Strikes & Error \\
    \hline
    4   & 29 & 16.8       \\
    10  & 24 & 4.87       \\
    13  & 20 & 4.95       \\
    13  & 9  & 0.146      \\
    17  & 12 & 0.302      \\
    32  & 36 & 2.11       \\
    60  & 19 & 1.09       \\
    88  & 22 & 1.78       \\
    88  & 11 & 1.65       \\
    186 & 7  & 4.79E-07   \\
    277 & 11 & 0.103      \\
    368 & 8  & 2.58       \\
    \hline
  \end{tabular}
\end{table}
The accuracy of price estimation is very good, except for the first
block, where, however, options have a very short time to
maturity. Recall that, as already remarked, in this case the algorithm
is induced to approximate the Heston density as closely as possible
only on \(\mathopen]-\infty,\log K_\mathrm{max}\mathclose]\), where
\(K_\mathrm{max}\) stands for the highest strike price
observed.

\subsection{Out-of-sample performance on real data}
We are now going to discuss the results of an extensive empirical
study on the out-of-sample performance of the option price estimators
discussed so far. The data set and the framework are the same used in
\cite{cm:Herm}, i.e. the data are raw option prices on the S\&P500 for
the whole year 2012, as mentioned above. For each trading day options
are divided in blocks with the same time to maturity, and on each
block instances of the parametric and nonparametric estimators of
option prices discussed above are calibrated to all prices bar one,
which is used as test price \(\pi\). The prices estimated by the
various calibrated parameters are then compared to the test price, and
absolute relative errors are computed. That is, if \(\hat{\pi}\) is an
estimated price, the error \(\abs{\hat{\pi}/\pi - 1}\) is
recorded. This operation is repeated until all prices in the block
have been used as test prices. The same procedure is applied to every
block of every day, obtaining samples of sizes between 40\,000 and
43\,500. Points of the empirical distribution of the absolute relative
pricing errors are collected in Tables \ref{tab:BS-Hes-VG} and
\ref{tab:Hermite}. In Table \ref{tab:BS-Hes-VG} we have also included,
as terms of comparison, the simplest parametric model, that is the
Black-Scholes model, and the simplest nonparametric model, that is the
Black-Scholes model with interpolation on the implied volatility
curve.
\begin{table}[tbp]
  \centering
  \caption{Pricing errors of Black-Scholes, variance-gamma, and Heston
    estimators}

  \vspace{2mm}

  \parbox{\textwidth}{\footnotesize The table reports selected
    quantiles of the empirical distribution of out-of-sample absolute
    pricing errors in percentage points of the Black-Scholes,
    variance-gamma, and Heston estimators. Figures in parentheses
    correspond to options with strike prices within the range of strike
    prices of options used to calibrate.}

  \vspace{4mm}
  
  \footnotesize{{\begin{center} 
 \begin{tabular*}{\textwidth}{@{\extracolsep{\fill}}lccccccc} 
 \toprule 
 \multicolumn{6}{c}{}\\ 
 \multicolumn{6}{c}{\textsc{Empirical distribution of Pricing errors}}\\ 
 \multicolumn{6}{c}{}\\ 
 \multicolumn{6}{c}{Black \& Scholes}\\ 
 \midrule 
Quantiles & $n=1$ &$n=2$ & $n=3$ & $n=4$ & $n=5$ \\  
 \cmidrule{2-6} 
$10\%$     & 10.5 (11.6)& 10.4 (11.6)& 10.4 (11.6)& 10.4 (11.8)& 10.5 (12.0)\\ 
$25\%$     & 21.9 (22.6)& 21.8 (22.5)& 21.8 (22.6)& 21.8 (22.7)& 21.9 (22.8)\\ 
$50\%$     & 38.0 (38.2)& 37.6 (38.1)& 37.4 (38.1)& 37.3 (38.1)& 37.2 (38.2)\\ 
$75\%$     & 80.0 (80.0)& 80.0 (80.0)& 80.0 (80.0)& 80.0 (80.0)& 80.0 (80.0)\\ 
$90\%$     & 93.5 (93.0)& 93.3 (92.9)& 93.3 (92.9)& 93.3 (92.9)& 93.3 (92.9)\\ 
$95\%$     & 96.4 (96.0)& 96.3 (96.0)& 96.3 (96.0)& 96.2 (95.9)& 96.1 (95.9)\\ 
 \multicolumn{6}{c}{}\\ 
 \multicolumn{6}{c}{Black \& Scholes with interpolated implied volatility}\\ 
 \midrule 
Quantiles & $n=1$ &$n=2$ & $n=3$ & $n=4$ & $n=5$ \\  
 \cmidrule{2-6} 
$10\%$     & 0.1 (0.1)& 0.1 (0.1)& 0.1 (0.1)& 0.1 (0.1)& 0.1 (0.1)\\ 
$25\%$     & 0.3 (0.2)& 0.3 (0.2)& 0.3 (0.2)& 0.3 (0.3)& 0.3 (0.3)\\ 
$50\%$     & 1.4 (1.0)& 1.3 (1.0)& 1.3 (1.0)& 1.3 (1.0)& 1.3 (1.1)\\ 
$75\%$     & 7.8 (4.5)& 7.4 (4.5)& 7.1 (4.5)& 6.9 (4.6)& 6.8 (4.7)\\ 
$90\%$     & 30.3 (14.2)& 28.7 (14.1)& 27.3 (14.2)& 26.4 (14.3)& 25.6 (14.4)\\ 
$95\%$     & 55.5 (25.7)& 52.0 (25.6)& 49.7 (25.7)& 48.2 (25.8)& 46.7 (25.9)\\ 
 \multicolumn{6}{c}{}\\ 
 \multicolumn{6}{c}{Variance-gamma}\\ 
 \midrule 
Quantiles & $n=1$ &$n=2$ & $n=3$ & $n=4$ & $n=5$ \\  
 \cmidrule{2-6} 
$10\%$     & 0.7 (0.8)& 0.7 (0.8)& 0.7 (0.8)& 0.7 (0.8)& 0.8 (0.8)\\ 
$25\%$     & 2.0 (2.0)& 2.0 (2.0)& 2.1 (2.0)& 2.1 (2.1)& 2.2 (2.1)\\ 
$50\%$     & 6.1 (5.7)& 6.0 (5.7)& 6.1 (5.8)& 6.1 (5.8)& 6.2 (5.9)\\ 
$75\%$     & 16.3 (15.1)& 16.2 (15.1)& 16.3 (15.2)& 16.3 (15.3)& 16.4 (15.4)\\ 
$90\%$     & 35.5 (32.3)& 35.2 (32.3)& 35.1 (32.4)& 35.0 (32.4)& 35.0 (32.5)\\ 
$95\%$     & 50.3 (46.7)& 50.0 (46.6)& 50.0 (46.7)& 50.0 (46.6)& 50.0 (46.7)\\ 
 \multicolumn{6}{c}{}\\ 
 \multicolumn{6}{c}{Heston}\\ 
 \midrule 
Quantiles & $n=1$ &$n=2$ & $n=3$ & $n=4$ & $n=5$ \\  
 \cmidrule{2-6} 
$10\%$     & 0.4 (0.4)& 0.4 (0.4)& 0.4 (0.4)& 0.4 (0.4)& 0.4 (0.4)\\ 
$25\%$     & 1.2 (1.2)& 1.2 (1.2)& 1.2 (1.2)& 1.2 (1.2)& 1.2 (1.2)\\ 
$50\%$     & 4.0 (3.8)& 4.0 (3.8)& 4.0 (3.8)& 4.0 (3.8)& 4.1 (3.9)\\ 
$75\%$     & 13.2 (12.4)& 13.2 (12.4)& 13.2 (12.4)& 13.2 (12.5)& 13.3 (12.6)\\ 
$90\%$     & 34.1 (29.4)& 33.8 (29.4)& 33.6 (29.4)& 33.4 (29.4)& 33.1 (29.4)\\ 
$95\%$     & 56.8 (45.7)& 56.1 (45.7)& 55.4 (45.6)& 54.6 (45.6)& 53.9 (45.5)\\ 
\midrule 
Test points& 43469 (37760)& 42755 (37522)& 41815 (37052)& 40830 (36461)& 39834 (35797)\\ 
\bottomrule 
\end{tabular*} 
\end{center}}}

  \label{tab:BS-Hes-VG}
\end{table}

\begin{table}[tbp]
  \centering
  \caption{Pricing errors of Hermite estimators}

  \vspace{2mm}

  \parbox{\textwidth}{\footnotesize The table reports selected
    quantiles of the empirical distribution of out-of-sample absolute
    pricing errors in percentage points of the Hermite estimators
    \(f^\ast_{n,p}\), \(f^\ast_n\), and \(f^\ast_{n,m}\).}

  \vspace{4mm}

  \footnotesize{{\begin{center} 
 \begin{tabular*}{\textwidth}{@{\extracolsep{\fill}}lccccccc} 
 \toprule 
 \multicolumn{6}{c}{}\\ 
 \multicolumn{6}{c}{\textsc{Empirical distribution of Pricing errors}}\\ 
 \multicolumn{6}{c}{}\\ 
 \multicolumn{6}{c}{$f^\ast_{n,p}$}\\ 
 \midrule 
Quantiles & $n=1$ &$n=2$ & $n=3$ & $n=4$ & $n=5$ \\  
 \cmidrule{2-6} 
$10\%$     & 2.9 (3.2)& 0.4 (0.4)& 0.4 (0.5)& 0.1 (0.1)& 0.1 (0.2)\\ 
$25\%$     & 8.8 (8.6)& 1.5 (1.5)& 1.6 (1.6)& 0.5 (0.5)& 0.5 (0.5)\\ 
$50\%$     & 19.4 (18.3)& 4.6 (4.4)& 5.3 (5.0)& 1.8 (1.7)& 1.8 (1.7)\\ 
$75\%$     & 37.2 (33.3)& 12.2 (11.0)& 13.3 (12.0)& 6.6 (5.8)& 6.5 (5.7)\\ 
$90\%$     & 66.7 (60.8)& 27.9 (23.5)& 28.3 (23.5)& 21.0 (16.3)& 21.7 (16.5)\\ 
$95\%$     & 80.0 (74.8)& 47.5 (36.2)& 46.5 (35.3)& 44.9 (28.7)& 50.3 (30.5)\\ 
 \multicolumn{6}{c}{}\\ 
 \multicolumn{6}{c}{$f^\ast_n$}\\ 
 \midrule 
Quantiles & $n=1$ &$n=2$ & $n=3$ & $n=4$ & $n=5$ \\  
 \cmidrule{2-6} 
$10\%$     & 1.3 (1.4)& 0.1 (0.2)& 0.2 (0.2)& 0.1 (0.1)& 0.1 (0.1)\\ 
$25\%$     & 4.2 (4.2)& 0.5 (0.5)& 0.6 (0.5)& 0.4 (0.4)& 0.4 (0.4)\\ 
$50\%$     & 10.5 (10.1)& 1.9 (1.8)& 2.0 (1.9)& 1.6 (1.5)& 1.6 (1.5)\\ 
$75\%$     & 22.7 (21.0)& 6.8 (6.0)& 6.8 (6.1)& 6.4 (5.5)& 6.5 (5.6)\\ 
$90\%$     & 45.9 (40.5)& 19.7 (15.9)& 19.5 (15.8)& 22.4 (16.4)& 23.7 (17.3)\\ 
$95\%$     & 63.1 (56.5)& 37.4 (27.0)& 38.0 (27.0)& 51.3 (30.8)& 59.8 (34.1)\\ 
 \multicolumn{6}{c}{}\\ 
 \multicolumn{6}{c}{$f^\ast_{n,m}$}\\ 
 \midrule 
Quantiles & $n=1$ &$n=2$ & $n=3$ & $n=4$ & $n=5$ \\  
 \cmidrule{2-6} 
$10\%$     & 9.6 (10.8)& 10.1 (11.3)& 0.8 (0.9)& 0.6 (0.7)& 0.5 (0.6)\\ 
$25\%$     & 20.6 (21.5)& 23.5 (24.4)& 2.5 (2.6)& 2.2 (2.3)& 1.8 (1.8)\\ 
$50\%$     & 36.0 (36.8)& 38.7 (39.3)& 6.2 (6.3)& 6.7 (6.7)& 5.0 (4.7)\\ 
$75\%$     & 79.0 (79.3)& 75.0 (75.0)& 15.1 (14.6)& 15.7 (15.2)& 11.8 (10.5)\\ 
$90\%$     & 93.3 (92.9)& 92.3 (91.7)& 45.1 (37.0)& 34.5 (31.4)& 24.9 (20.4)\\ 
$95\%$     & 96.3 (96.0)& 95.7 (95.3)& 66.7 (64.5)& 56.4 (50.0)& 39.5 (30.9)\\ 
\midrule 
Test points& 43469 (37760)& 42755 (37522)& 41815 (37052)& 40830 (36461)& 39834 (35797)\\ 
\bottomrule 
\end{tabular*} 
\end{center}}}

  \label{tab:Hermite}
\end{table}
Each panel of Table \ref{tab:BS-Hes-VG} contains five columns indexed
by different values on \(n\), where \(n=k\), \(k=1,\ldots,5\) means
that all blocks in the data set containing at most \(k\) options have
been excluded from the computations. This explains the different
sample sizes reported in the last row of the table. The rationale for
this is that a model with \(k\) parameters calibrated to \(k\) or less
prices obviously suffers of the well-known problem of overfitting.
On the other hand, the empirical results show that the distributions
of errors of all parametric models are quite robust with respect to
the presence of small blocks, with minor variations showing up mostly
at the 90\% and 95\% quantiles.
The columns of Table \ref{tab:Hermite} have the same interpretation,
but in this case, if \(n=k\), results are reported for estimators
\(f^\ast_{k,p}\), \(f^\ast_k\), and \(f^\ast_{k,m}\) in the first,
second, and third panel, respectively.

Inspection of Table \ref{tab:BS-Hes-VG} reveals that both parametric
estimators, i.e. the variance-gamma and the Heston model, are major
improvements with respect to the elementary Black-Scholes
model.\footnote{On the other hand, the Black-Scholes model with
  interpolation on the implied volatility still has lower pricing
  errors than the two parametric methods. This is however not strictly
  relevant for the issues investigated here, as interpolating on a
  volatility curve is not a method depending on a fixed number of
  parameters.}  Unsurprisingly, the five-parameter Heston model
  displays lower average errors (up to the 90\% quantile) than the
  three-parameter variance-gamma model. In particular, the median
  absolute relative error of the Heston model is 4\%, while the
  corresponding error of the variance-gamma model is around 50\%
  higher.
For a detailed discussion of the results concerning the Hermite
estimators (including others not considered here), we refer to
\cite{cm:Herm}.

In view of the discussion in \S\S\ref{ssec:VGtoH}-\ref{ssec:HestoH},
we focus on the comparison between the variance-gamma model and the
Hermite models with three parameters, that are \(f^\ast_{1,p}\) and
\(f^\ast_{3,m}\), as well as between the Heston model and the Hermite
models with five parameters, that are \(f^\ast_{3,p}\), \(f^\ast_2\),
and \(f^\ast_{5,m}\).
The variance-gamma price estimator performs clearly better than the
Hermite \(f^\ast_{1,p}\) estimator, with the median error of the
latter more than three times as large as the median error of the
former, and the empirical error distribution of the latter dominating
the one of the former. The picture is not as clear when the
variance-gamma estimator is compared to the Hermite \(f^\ast_{3,m}\)
estimator: the quantiles of their respective empirical error
distributions are very close up to and including the median, but the
75\% quantile corresponding to the Hermite estimator is 15.1\%, hence
better than the same quantile corresponding to the variance-gamma,
that is 16.3\%. On the other hand, large estimation errors are less
frequent for the variance-gamma than for the Hermite estimator.
The comparison between the five-parameter estimators reveals some
interesting and perhaps unexpected phenomena. Perhaps most
importantly, it turns out that the empirical error of the Hermite
estimator \(f^\ast_2\) is much lower, at all quantile levels, than the
empirical error of the Heston model. This is indeed quite surprising,
as the Heston model is tailor-made for financial returns, while the
Hermite estimator is completely ``agnostic''. Also the Hermite
estimators \(f^\ast_{3,p}\) and \(f^\ast_{5,m}\) do not perform badly
compared to the Heston estimator: both Hermite estimators have very
close error up to the median, that are nonetheless higher than the
errors of the Heston estimator. On the other hand, at higher
quantiles, the Heston estimator performs worse than \(f^\ast_{3,p}\),
which in turns performs worse than \(f^\ast_{5,m}\).


\section{Concluding remarks}
The empirical performance of both parametric and nonparametric
estimators of option prices, focusing on variance-gamma and Heston
models for the former and on expansions in Hermite functions for the
latter, was evaluated by the off-sample relative pricing error using
historical data on European options on the S\&P500 index. 

The main empirical observation is that a simple Hermite estimator with
the same number of parameters as the Heston model outperforms the
latter. This is remarkable as Hermite estimators are ``uninformed'' by
their nature, while the Heston model is designed for financial assets.
Moreover, as expected, the variance-gamma model, that depends on less
parameters, has larger average pricing errors than the Heston model.
These results highlight the trade-offs between the flexibility of
nonparametric methods and the domain-specific precision of parametric
models. Notably, the performance of Hermite estimators, combined with
their simple and fast calibration, makes them a valuable tool for
validation purposes.

A relevant question, that would be natural to address in future
contributions, is whether suitable time-dependent extensions of
Hermite estimators can still keep up with Heston estimators in the
pricing of options with different times to maturity.


\appendix
\section{Computational issues}
\label{app:num}
We collect here, for completeness, some remarks about practical
computational issues.
All numerical computations have been done with
\href{https://www.octave.org}{Octave} 9.1.0 on Linux, using
\texttt{fminsearch} as minimization algorithm.
The minimization problem to which calibration of a model amounts is
often marred by the issue of local minima, and the cases of
variance-gamma and Heston densities are not exception. In the
variance-gamma case, calibration uses as starting point of the
optimization the point \((\theta_0,\sigma_0,\alpha_0) =
(0.1,0.3,2)\). Moreover, in order to avoid degenerate situations, a
lower bound on \(\sigma\) equal to \(0.05\) was enforced.
To price (European put) options in the Heston model, hence also for
calibration purposes, we used the code provided by \cite{Frei:Heston},
that implements the steps described in \S\ref{sec:Heston}. The
starting point of calibrations is
\((v_{00},\theta_0,\rho_0,\kappa_0,\eta_0) = (0.02, 0.35, -0.5, 0.5,
0.3)\). On the other hand, Heston densities are approximated by an FFT
implementation of Fourier inversion of the characteristic function,
using code provided by \cite{GG:Hes}.


\bibliographystyle{plainurl}
\bibliography{ref,finanza}

\end{document}